\documentclass[12pt]{article}
\usepackage{graphicx}
\usepackage{epstopdf}
\usepackage{amssymb}
\usepackage{amsmath}
\begin{document}
\baselineskip 0.6cm
\newcommand{\tri}{\triangleright}
\newcommand{\range}{{\rm range}}
\newcommand{\Ree}{{\rm Re }}
\newcommand{\Imm}{{\rm Im }}
\newcommand{\diag}{{\rm diag}}
\newcommand{\sign}{{\rm sign}}
\newcommand{\tr}{{\rm tr}}
\newcommand{\rank}{{\rm rank}}
\newcommand{\bp}{\bigskip}
\newcommand{\mdp}{\medskip}
\newcommand{\slp}{\smallskip}
\newcommand{\Rw}{\Rightarrow}
\newcommand{\ts}{& \hspace{-0.1in}}
\newcommand{\nn}{\nonumber}
\newcommand{\bea}{\begin{eqnarray}}
\newcommand{\eea}{\end{eqnarray}}
\newcommand{\beas}{\begin{eqnarray*}}
\newcommand{\eeas}{\end{eqnarray*}}
\newcommand{\beq}{\begin{equation}}
\newcommand{\eeq}{\end{equation}}
\newtheorem{exa}{Example}[section]
\newtheorem{thm}{Theorem}[section]
\newtheorem{lem}{Lemma}[section]
\newtheorem{prop}{Proposition}[section]
\newtheorem{fact}{Fact}[section]
\newtheorem{cor}{Corollary}[section]
\newtheorem{defn}{Definition}[section]
\newtheorem{rem}{Remark}[section]
\renewcommand{\theequation}{\thesection.\arabic{equation}}

\title{{\bf Adaptive Output Tracking Control with Reference Model
    System Uncertainties}}
\author{{\it Gang Tao}\\
\normalsize Department  of Electrical and Computer Engineering\\  
\normalsize University of Virginia \\ 
\normalsize Charlottesville, VA 22903, USA} 
\date{} 
\maketitle 
\begin{abstract}
This paper develops adaptive output tracking control schemes with the
reference output signal generated from an unknown reference system
whose output derivatives are also unknown. To deal with such reference
system uncertainties, an expanded adaptive controller structure is
developed to include a parametrized estimator of the equivalent
reference input signal. Without using the knowledge of the reference system
transfer function and equivalent input, both are the critical
components of a traditional model reference adaptive control (MRAC)
scheme, the developed new MRAC schemes designed for various cases plant and
reference model uncertainties, ensure completely parametrized error
systems and stable parameter adaptation, leading to the desired
closed-loop system stability and asymptotic output tracking.
\end{abstract}

\bigskip
 {\bf Keywords}: 
Adaptive control, 
output tracking, 
parameter uncertainties, 
plant-model matching, 
reference model system. 

\setcounter{equation}{0}
\section{Problem Description}
Consider a linear time-invariant (LTI) system (plant) described by
\beq
\dot{x}(t) = A x(t) + b u(t),\;y(t) = c x(t),
\label{5x1}
\eeq
where $x(t) \in R^{n}$, $u(t) \in R$ and $y(t) \in R$ are the state
vector, input and output signals, and $A \in R^{n \times n}$, $b \in
R^{n}$ and $c \in R^{1 \times n}$ are unknown constant matrices such
that $G(s) = c(sI - A)^{-1}b$ has relative degree $n^*$. 
The control objective is to design the input signal $u(t)$
to ensure the closed-loop system stability and asymptotic output
tracking: $\lim_{t \rightarrow \infty}(y(t) - y_m(t)) = 0$, where
$y_m(t)$ is the output of a stable reference model system:
\beq
\dot{x}_{m}(t) = A_{m} x_{m}(t) + b_{m} u_m(t),\;y_m(t) = c_m x_m(t)
\label{5x0}
\eeq
where $x_{m}(t) \in R^{n}$, $u_m(t) \in R$ and $y_m(t) \in R$ are
the reference system state vector, input and output signals, and
$A_{m} \in R^{n \times n}$, $b_{m} \in R^{n}$ and $c_m \in R^{1 \times
  n}$ are some constant matrices such that $G_m(s) = c_m(sI -
A_m)^{-1} b_m$ has relative degree $n_m^*$.

\bigskip
{\bf The traditional model reference adaptive control
 problem}. The traditional adaptive output tracking control
problem (the model reference adaptive control (MRAC) 
problem) is to achieve the control objective under the condition that $G_m(s) = c_m(sI -
A_m)^{-1} b_m$ is known and $n_m^* \geq n^*$. The reference model
system is typically expressed in an input-output system form
\beq
y_m(t) = W_m(s)[r_m](t),
\label{5mrs}
\eeq
where $W_m(s) = \frac{1}{P_m(s)}$ for a chosen stable polynomial
$P_m(s)$ of degree $n^*$ (as a convenient notation, $y_m(t) = W_m(s)[r](t)$
denotes the output $y_m(t)$ of a system $W_m(s)$ with input
$r_m(t)$), and $r_m(t)$ is a given (available) reference input. In
terms of $G_m(s)$ and $u_m(s)$, we have $y_m(t) =
G_m(s)[u_m](t) = W_m(s) P_m(s) G_m(s)[u_m](t)$, so that $r_m(t) = P_m(s)
G_m(s)[u_m](t)$ which is known and available from $u_m(t)$. 
Note that both the reference system transfer function
$W_m(s)$ and the reference input $r_m(t)$ are both used in a MRAC scheme.

On the other hand, if $y_m(t)$ and its time-derivatives $y_m^{(i)}(t)$, $i=1,2,\ldots,
n^*$, are given, then, one can choose a stable $P_m(s)$ and form 
$r_m(t) = P_m(s)[y_m](t)$, for adaptive control design.

\bigskip
{\bf The current adaptive control problem}. The problem to be solved
in this paper is adaptive output tracking control without the
knowledge of the reference system $(A_m, b_m, c_m)$ or $G_m(s) =
c_m(sI - A_m)^{-1} b_m$ and that of the time-derivatives
$y_m^{(i)}(t)$, $i=1,2,\ldots, n^*$. This is an open adaptive output
tracking control problem with an uncertain reference model system.

Without such reference system knowledge, the
original reference model system (\ref{5x0}): $y_m(t) = G_m(s)[u_m](t)$,
cannot be expressed in the form (\ref{5mrs}): 
$y_m(t) = W_m(s)[r_m](t)$ with an available $r_m(t)$, as 
$r_m(t) = P_m(s) G_m(s)[u_m](t) = P_m(s)[y_m](t)$ depends on either the
unknown $G_m(s)$ or the unknown time-derivatives $y_m^{(i)}(t)$,
$i=1,2,\ldots, n^*$. Thus, the traditional MRAC solution is not
applicable to the current adaptive control problem studied in this
paper. 

In this case, the reference system (\ref{5mrs}): $y_m(t) =
W_m(s)[r_m](t)$, is only a virtual reference model system as $r_m(t)$
is unknown but $y_m(t)$ is known from the actual reference system
(\ref{5x0}).

\bigskip
{\bf A leader-follower tracking problem}. The above adaptive output
tracking control problem can be considered as a leader-follower
tracking problem in which the reference model system (the leader)
parameters $(A_m, b_m, c_m)$ or transfer function $G_m(s) = c_m(sI -
A_m)^{-1} b_m$ and the time-derivatives $y_m^{(i)}(t)$, $i=1,2,\ldots,
n^*$, of the leader output $y_m(t)$ are unknown. The follower
is the controlled system (plant) (\ref{5x1}) whose output $y(t)$ is to
track the leader output $y_m(t)$. Because of the leader (reference
system) uncertainties, such a problem cannot be solved by a
traditional MRAC solution which requires the knowledge of $G_m(s)$ or 
$y_m^{(i)}(t)$, $i=1,2,\ldots, n^*$, as explained above. 

New solutions are needed, to be developed in this paper, for such a
new adaptive control problem, based on expanded controller
structures using parametrizations of $r_m(t) = P_m(s)[y_m](t)$.

\setcounter{equation}{0}
\section{Technical Background}
\label{Technical Background}
In this section, we present some background of model reference
adaptive control, summarize the paper's technical contributions,
describe its organization, and review some related literature.

\bigskip
{\bf MRAC background}. As developed in \cite{is96}, \cite{na89},
\cite{sb89}, \cite{t03}, with (\ref{5mrs}) being the chosen reference
model system for $r_m(t) = r(t)$ being a chosen reference input
signal, the traditional output feedback (reduced-order observer based)
model reference adaptive controller structure is
\begin{equation}
u(t) = \theta_{1}^{T}(t) \omega_{1}(t) + \theta_{2}^{T}(t)
\omega_{2}(t) + \theta_{20}(t) y(t) + \theta_{3}(t) r(t),
\label{5732A.930}
\end{equation}
where $\theta_{1} \in R^{n-1}$, $\theta_{2} \in R^{n-1}$, $\theta_{20}
\in R$ and $\theta_{3} \in R$, and 
\begin{equation}
\omega_{1}(t) = \frac{a(s)}{\Lambda(s)}[u](t),\;\omega_{2}(t) = 
\frac{a(s)}{\Lambda(s)}[y](t)
\end{equation}
with $a(s) = [1,s,\cdots,s^{n-2}]^{T}$ and $\Lambda(s)$ being a monic
stable polynomial of degree $n-1$. 
 The parameters $\theta_{1}(t) \in R^{n-1}$, $\theta_{2}(t) \in
 R^{n-1}$, $\theta_{20}(t) \in R$ and $\theta_{3}(t) \in R$ are the
 adaptive estimates (to be updated from an adaptive law) of the constant parameters
$\theta_{1}^{*} \in R^{n-1}$, $\theta_{2}^{*} \in R^{n-1}$, 
$\theta_{20}^{*} \in R$  and $\theta_3^* = \frac{1}{k_p}$, which
 satisfy the polynomial matching equation:
\bea
\ts \ts \theta_{1}^{*T} a(s) P(s) 
+ (\theta_{2}^{*T} a(s) + \theta_{20}^{*} 
\Lambda(s)) k_{p} Z(s)\nn\\
\ts  =\ts \Lambda(s)(P(s) - k_{p}  
\theta_{3}^{*} Z(s) P_{m}(s)),
\label{5732A.951}
\eea
for $G(s) = c(sI - A)^{-1} b = k_p \frac{Z(s)}{P(s)}$ with monic
polynomials $Z(s)$ and $P(s)$ of degrees $n-n^*$ and $n$
respectively. It can be shown that the tracking error $e(t) = y(t) -
y_m(t)$ satisfies
\beq
e(t) = k_p W_m(s)[(\theta_1 - \theta_1^*)^T \omega_1 + 
(\theta_2 - \theta_2^*)^T \omega_2 + (\theta_{20} - \theta_{20}^*) y +
(\theta_3 - \theta_3^*) r](t).
\label{5tee0}
\eeq
Based on this error equation, an adaptive scheme can be designed to
achieve the control objective.

\medskip
As given in \cite{t03}, the traditional state feedback model
reference adaptive controller structure is 
\beq
u(t) = k_1^T(t) x(t) + k_2(t) r(t),
\label{5sfcs}
\eeq
where $k_1(t)$ and $k_2 (t) \in R$ are the adaptive estimates of the
unknown parameters $k_1^* \in R^n$ and $k_2^* \in R$ which satisfy the
matching equation
\beq
c(sI - A - b k_1^{*T})^{-1} b k_2^* = W_m(s).
\label{5me2}
\eeq
Similar to (\ref{5tee0}), in this case, the tracking error $e(t) = y(t) - y_m(t)$ satisfies
\beq
e(t) = k_p W_m(s)[(k_1 - k_1^*)^T x + (k_2 - k_2^*) r](t).
\label{5tee00}
\eeq
 
Under the availability of the state vector signal $x(t)$, a state
feedback controller (\ref{5sfcs}) has a less complex adaptive control
scheme than an output feedback controller (\ref{5732A.930}) does. Both
schemes also use the chosen reference system transfer function
$W_m(s)$ to generate auxiliary signals such as estimation errors and
regressors for designing the stable adaptive parameter update laws.

\medskip
As given in \cite{is96}, \cite{na89}, \cite{sb89}, \cite{t03}, the
basic model reference adaptive control (MRAC) assumptions for the system
(\ref{5x1}) with unknown parameters $(A, b, c)$ are

\begin{description}
\item[] {\bf Assumption (A1)}: $(A, b, c)$ is stabilizable and
  detectable, all zeros of $G(s) = c(sI - A)^{-1} b = k_p
  \frac{Z(s)}{P(s)}$ are stable, the relative degree $n^*$ of $G(s)$
  is known, and the sign $\sign[k_p]$ of $k_p$ is known. 
\end{description}

For the new adaptive output tracking control problem to be solved this
paper, the assumptions on the reference model system (\ref{5x0}): 
$\dot{x}_{m}(t) = A_{m} x_{m}(t) + b_{m} u_m(t),\;y_m(t) = c_m x_m(t)$, are

\begin{description}
\item[] {\bf Assumption (A2)}: The reference model system $(A_m, b_m, c_m)$
  is stabilizable and detectable, its signals $y_m(t)$, $x_m(t)$
  and $u_m(t)$ are bounded with $y_m(t)$ (and $x_m(t)$) and $u_m(t)$
  known, and its transfer function 
  $G_m(s) = c_m(sI - A_m)^{-1} b_m$ has relative degree
  $n_m^* \geq n^*$.
\end{description}

{\bf Contributions of this paper}. 
As clarified above, in view of (\ref{5x0}) and (\ref{5mrs}), the
equivalent signal $r_m(t) = P_m(s) G_m(s)[u_m](t) = P_m(s)[y_m](t)$, 
due to the uncertainties of $G_m(s)$ or $y_m^{(i)}(t)$,
$i=1,2,\ldots, n^*$, is not available in the current new adaptive
control problem for which the traditional controller structures
(\ref{5732A.930}) and (\ref{5sfcs}) are not applicable and need to be
redesigned to develop new solutions to the current adaptive control
problem, to deal with the reference system uncertainties.

\medskip
The contributions of this paper are to develop a new method of
parametrizing and estimating the unknown equivalent reference input
signal $r_m(t)$ for adaptive output tracking control in the presence
of reference model system uncertainties as described above, to
develop adaptive control schemes using either state feedback or output
feedback and using either the reference system state variables
$x_m(t)$ or the reference system output $y_m(t)$, for control design,
and to conduct a comprehensive study of the new adaptive control problem
and its new solutions, addressing their extensions.

\medskip
\medskip
{\bf Paper organization}. 
The paper is organized as follows. A literature review will be given
next. The state feedback adaptive control designs will be developed in
Section 3, and the output feedback adaptive control designs will be
developed in Section 4. For both sections, designs using $x_m(t)$ and
using $y_m(t)$ will be derived. All designs make use of the reference
system input signal $u_m(t)$, with parametrization and estimation of
the unavailable signal $r_m(t)$, as the new features of the developed
adaptive control schemes for solving the current new adaptive control
problem. Desired adaptive control system properties are analyzed in
Section 5. Simulation results illustrating the desired system
performance are presented in Section 6.

\bigskip
{\bf Literature review}. From the early work of output tracking model
reference adaptive control \cite{m74}, \cite{m80}, \cite{nlv80},
\cite{nv78}, a reference model system is designed (with a known
input $r(t)$, a known output $y_m(t)$ and a known transfer function
$W_m(s)$) before an adaptive controller is designed for an unknown
system (plant) whose transfer function has the same relative degree
$n^*$ as that of $W_m(s)$. Such a framework has been used since then
for the development of various extended adaptive control systems whose
theory and techniques have been summarized in the classical books
\cite{is96}, \cite{na89}, \cite{sb89} as well as \cite{kkk95} (where the
time-derivatives $y_m^{(i)}(t)$ of $y_m(t)$, $i=1,2,\ldots, n^*$, are
assumed to be known for a backstepping design of output tracking
control schemes, including that for linear systems). 

Adaptive tracking control using such full knowledge of reference model
systems has been extensively studied for various problems and has made
a vast amount of significant contributions as reported in the
literature, for example, \cite{af21} (historical perspective of adaptive control and
learning), \cite{ako08} (nonlinear and adaptive control with applications),
\cite{aw95} (adaptive control theory and applications),
\cite{ew82} (multivariable model reference adaptive control),
\cite{gs84} (discrete-time adaptive control), \cite{if06} (adaptive
control tutorial), \cite{kbs98} (direct adaptive control theory and
applications), \cite{llmk11} (adaptive control algorithms, analysis
and applications), \cite{lw13} (robust and adaptive control 
with aerospace applications), \cite{lq02} (adaptive control of
nonlinearly parameterized systems), \cite{oha03} (multivariable adaptive
control), \cite{smop02} (adaptive control and estimation using neural
and fuzzy approximator techniques), \cite{ot89}
(survey of robustness of adaptive controllers), \cite{t14} (survey
of multivariable adaptive control), \cite{tk96} (adaptive control of systems with
actuator and sensor nonlinearities), \cite{wwg20} (multivariable
adaptive control), \cite{wwz17} (adaptive control of systems with
actuator failures, subsystem interactions, and nonsmooth
nonlinearities), \cite{ysb16} (adaptive tracking control of
switched linear systems), \cite{zsc23} (adaptive
disturbance rejection control using partial-state feedback),
\cite{zw08} (adaptive control of systems with nonsmooth nonlinearities, interactions or
  time-variations). 

\medskip
The new adaptive output tracking control problem to be studied in this
paper is that the reference model system is not a designed all-known
system but is a given system whose input and output (and state
variables) are known but its transfer function $G_m(s)$ is unknown,
and so are the output time-derivatives $y_m^{(i)}(t)$, $i=1,2,\ldots,
n^*$, to the controller for the plant. Such an adaptive control problem is
practical in the sense that the controlled plant output is expected to
tracking the output of a given but uncertain dynamic system. Since the
reference system transfer function $G_m(s)$ and output
time-derivatives $y_m^{(i)}(t)$, $i=1,2,\ldots, n^*$, are unknown, the
traditional model adaptive control solutions are not applicable to
such a new adaptive tracking control problem. This goal of this paper
is to develop new adaptive controller structures to deal with the
reference model system uncertainties, to solve the new adaptive
control problem for various cases.

\setcounter{equation}{0}
\section{State Feedback Control Designs}
\label{State Feedback Control Designs}
We first develop new adaptive state feedback control schemes, assuming
that the state vector $x(t)$ of the controlled system (\ref{5x1}) is
available, to solve the new adaptive output tracking problem.

\subsection{Design Using Reference System State $x_m(t)$}
\label{Design Using Reference System State xm(t)}
Recall from (\ref{5x0}) and (\ref{5mrs}) that the equivalent signal
$r_m(t) = P_m(s) G_m(s)[u_m](t) = P_m(s)[y_m](t)$, used in the
traditional adaptive controller structures (\ref{5732A.930}) and
(\ref{5sfcs}), is not available in the current problem, due to the 
 uncertainties of $G_m(s)$ or $y_m^{(i)}(t)$, $i=1,2,\ldots, n^*$. A
 key solution step is to parametrize the unknown signal $r_m(t) =
P_m(s)[y_m](t)$. 

\medskip
Consider the reference system (\ref{5x0}): 
$\dot{x}_{m}(t) = A_{m} x_{m}(t) + b_{m} u_m(t),\;y_m(t) = c_m
x_m(t)$, with relative degree $n_m^* \geq n^*$. It follows that 
$c_m A_m^i b_m = 0$ for $i= 0,1,\ldots, n_m^*-2$, and $c_m A_m^{n_m^*-1} b_m
\not = 0$. We can then express the $i$th order time-derivative
$y_m^{(i)}(t) = \frac{d^i}{d t^i} y_m(t)$ of $y_m(t)$ as
\beq
y_m^{(i)}(t) = \left\{ \begin{array}{ll}
c_m A_m^{i} x_m(t) & \mbox{for $i=0,1,\ldots,n_m^* - 1$} \\ 
c_m A_m^{i} x_m(t) + c_m A_m^{i-1} b_m u_m(t) & \mbox{for $i=n_m^*$.}
\end{array}
\right.
\eeq

\medskip
{\bf Parametrization of $r_m(t)$ when $n_m^* = n^*$}. In this case, we
have
\beq
y_m^{(i)}(t) = \left\{ \begin{array}{ll}
c_m A_m^{i} x_m(t) & \mbox{for $i=0,1,\ldots,n^* - 1$} \\ 
c_m A_m^{i} x_m(t) + c_m A_m^{i-1} b_m u_m(t) & \mbox{for $i=n^*$.}
\end{array}
\right.
\eeq
Hence, for a chosen stable polynomial $P_m(s) = s^{n^*} + p_{n^*-1} s^{n^*-1}
+ \cdots + p_1 s + p_0$, we can express $r_m(t) = P_m(s)[y_m](t)$ as
\bea
\label{5r01}
r_m(t) \ts = \ts y_m^{(n^*)}(t) + p_{n^*-1} y_m^{(n^*-1)}(t) 
+ \cdots + y_m^{(1)}(t) + p_m y_m(t) \nn\\
\ts = \ts \alpha_1^T x_m(t) + \alpha_2 u_m(t),
\eea
where 
\bea
\label{5alpha1}
\alpha_1^T \ts = \ts c_m A_m^{n^*} + p_{n^*-1} c_m A_m^{n^*-1} + \cdots + p_1
c_m A_m + p_0 c_m \in R^{n \times 1}\\
\alpha_2 \ts = \ts c_m A_m^{n^*-1} b_m \in R.
\label{5alpha2}
\eea

\medskip
{\bf Parametrization of $r_m(t)$ when $n_m^* > n^*$}. In this case, 
we have $r_m(t) = P_m(s)[y_m](t)$ as
\bea
\label{5r04}
r_m(t) \ts = \ts y_m^{(n^*)}(t) + p_{n^*-1} y_m^{(n^*-1)}(t) 
+ \cdots + y_m^{(1)}(t) + p_m y_m(t) \nn\\
\ts = \ts \alpha_1^T x_m(t), 
\eea
where 
\beq
\alpha_1^T = c_m A_m^{n^*} + p_{n^*-1} c_m A_m^{n^*-1} + \cdots + p_1
c_m A_m + p_0 c_m \in R^{n \times 1}.
\eeq

\medskip
Note that a similar parametrization of $r_m(t)$ when $n_m^* < n^*$ would contain
the time-derivatives $u_m^{(i)}(t)$, $i=1,2,\ldots, n^*-n_m^*$, of
$u_m(t)$.

\medskip
\medskip
{\bf Nominal control law}. 
Based on the parametrized expressions (\ref{5r01}) and (\ref{5r04}) of
$r_m(t)$ for the cases of $n_m^* \geq
n^*$, the nominal state feedback control law is
\bea
u(t) \ts = \ts k_1^{*T} x(t) + k_2^* r_m(t) \nn \\
\ts = \ts  k_1^{*T} x(t) + 
k_{21}^{*T} x_m(t) + k_{22}^* u_m(t),
\label{5u1n}
\eea
where 
\beq
k_{21}^{*} = k_2^* \alpha_1,\;k_{22}^* = k_2^* \alpha_2,
\eeq
and $k_1^{*} \in R^n$ and $k_2^* \in R$ are constant
parameters satisfying (\ref{5me2}). 

It can then be shown that the closed-loop system output 
\beq
y(t) = c(sI - A - b k_1^{*T})^{-1} b k_2^*[r_m](t)
= W_m(s)[r_m](t) = y_m(t),
\label{5om}
\eeq
with the exponential decaying effect of the initial conditions
ignored, that is, more precisely, $\lim_{t \rightarrow \infty}(y(t) -
y_m(t)) = 0$ exponentially. 

\bigskip
{\bf Adaptive control law}. The adaptive state feedback control law is
\bea
u(t) \ts = \ts  k_1^{T} x(t) + 
k_{21}^{T} x_m(t) + k_{22} u_m(t),
\label{5u1a}
\eea
where $k_1 \in R^n$, $k_{21} \in R^n$ and $k_{22} \in R$ are the
estimates of $k_1^*$, $k_{21}^*$ and $k_{22}^*$,
and $x_m(t)$ and $u_m(t)$ are the reference system state and input signals.

\medskip
\medskip
{\bf Tracking error equation}. With (\ref{5u1a}), the closed-loop system with (\ref{5x1}) becomes
\bea
\dot{x}(t) \ts = \ts A x(t) + b (k_1^{T} x(t) + 
k_{21}^{T} x_m(t) + k_{22} u_m(t)) \nn\\
\ts = \ts A x(t) + b (k_1^{*T} x(t) + 
k_{21}^{*T} x_m(t) + k_{22}^* u_m(t)) \nn\\
\ts \ts + b ((k_1- k_1^*)^{T} x(t) + 
(k_{21}- k_{21}^*)^{T} x_m(t) + (k_{22}-k_{22}^*) u_m(t)).
\label{5x1c}
\eea

With (\ref{5om}), $k_2^* r_m(t) =
k_{21}^{*T} x_m(t) + k_{22}^* u_m(t)$ and $\rho^* = \frac{1}{k_2^*}$,
similar to (\ref{5tee00}), 
the tracking error $e(t) = y(t)- y_m(t)$ with $y(t) = c x(t)$ satisfies
\beq
e(t) = \rho^* W_m(s)[(k_1- k_1^*)^{T} x + 
(k_{21}- k_{21}^*)^{T} x_m + (k_{22}-k_{22}^*) u_m](t).
\label{5e(t)}
\eeq

\medskip
\medskip
{\bf Adaptive parameter update laws}. We introduce the
estimation error signal
\beq
\epsilon (t) =  e(t) + \rho(t) \xi(t),
\label{5732A.35}
\eeq
where $\rho(t)$ is the estimate of $\rho^{*}$, and 
\bea
\xi(t) \ts = \ts \theta^{T}(t) \zeta(t) - W_m (s) [\theta^{T} \omega](t) \in R\\
\theta (t) \ts = \ts \left[k_1^T(t), k_{21}^T(t), k_{22}(t)\right]^T \in R^{2n+1}\\
\omega (t) \ts = \ts \left[x^{T}(t), x_m^T(t), u_m(t) \right]^{T}\in R^{2n+1}\\
\zeta (t) \ts = \ts W_{m}(s)[\omega](t)\in R^{2n+1}.
\eea

With $\theta^{*} = \left[k_1^{*T}, k_{21}^{*T}, k_{22}^* \right]^T\in R^{2n+1}$, 
it can be verified that $\epsilon(t)$ has the linear form
\beq
\epsilon (t) = \rho^{*} (\theta(t) - \theta^{*})^{T} \zeta(t) + (\rho(t) - 
\rho^{*}) \xi(t).
\label{5732A.36}
\eeq

\medskip
The adaptive laws for $\theta(t)$ and $\rho(t)$ are chosen as
\bea
\label{5732A.37}
\dot{\theta} (t) \ts = \ts - \frac{\Gamma \sign[k_p] \zeta (t) \epsilon (t)} 
{m^2(t)} \\*[0.05in]
\dot{\rho} (t) \ts =  \ts- \frac{\gamma\, \xi(t) \epsilon (t)} 
{m^2(t)}
\label{5732A.38}
\eea
where $\Gamma = \Gamma^T > 0$ and $\gamma > 0$ are constant gains, 
and $m(t) = \sqrt{1 + \zeta^{T}(t) \zeta(t) + \xi^{2}(t)}$.

\medskip
With the new controller structure (\ref{5u1a}), this adaptive control
scheme, with the completely parametrized tracking error equation 
(\ref{5e(t)}) and estimation error equation (\ref{5732A.36}),
has the desired parameter adaptation properties (see Lemma
\ref{5lemma11}) and closed-loop system stability and asymptotic output
tracking ($\lim_{t \rightarrow \infty} (y(t) - y_m(t)) = 0$) 
properties (see Theorem \ref{5thm1}).

\subsection{Design Using Reference System Output $y_m(t)$}
For the reference system (\ref{5x0}): 
$\dot{x}_{m}(t) = A_{m} x_{m}(t) + b_{m} u_m(t),\;y_m(t) = c_m
x_m(t)$ with $x_m(t)$ unavailable, we can build a nominal reduced-order state
observer \cite[page 272]{r96}, \cite[page 172]{t03}, based on the
information of $y_m(t)$ and $u_m(t)$, to
generate an estimate $\hat{x}_{m}(t)$ of $x_m(t)$ such that 
$\lim_{t \rightarrow \infty} (\hat{x}_{m}(t) - x_m(t)) = 0$
exponentially. Such a reduced-order observer based estimate is
\beq
\hat{x}_m(t) = Q_m \left[\begin{array}{c}
y_m(t)\\
w_m(t) + L_r y_m(t)
\end{array}
\right],
\label{5hatxm}
\eeq
where $Q_m \in R^{n\times n}$ is a nonsingular matrix such that $y_m(t)
= [1, 0, \ldots, 0] Q_m^{-1} x_m(t)$, 
$L_r \in R^{n-1}$ is a reduced-order observer gain vector, and the
dynamic part $w_m(t)$ of $\hat{x}_m(t)$ is
\beq
w_m(t) = \frac{G_{u_m}(s)}{\Lambda_e(s)}[u_m](t) + 
 \frac{G_{y_m}(s)}{\Lambda_e(s)}[y_m](t) \in R^{n-1}
\label{5wm(t)}
\eeq
for a chosen monic stable polynomial $\Lambda_e(s)$ of degree
$n-1$, and some $(n-1)$-dimensional polynomial vectors $G_{u_m}(s)$
and $G_{y_m}(s)$.

Replacing $x_m(t)$ with $\hat{x}_m(t)$, we can express $r_m(t) = 
\alpha_1^T \hat{x}_m(t) + \alpha_2 u_m(t)$ as
\beq
r_m(t) = \beta_1^T \omega_{u_m}(t) + \beta_{2}^T \omega_{y_m}(t) +
\beta_{20} y_m(t) + \alpha_2 u_m(t)
\label{5r02}
\eeq
for some $\beta_1 \in R^{n-1}$, $\beta_{2} \in R^{n-1}$ and
$\beta_{20} \in R$, where, for $a(s) = [1, s, \ldots, s^{n - 2}]^{T}$, 
\beq
\omega_{u_m}(t) = \frac{a(s)}{\Lambda_e(s)}
[u_m](t),\;\omega_{y_m}(t) = \frac{a(s)}{\Lambda_e(s)}[y_m](t).
\eeq

\bigskip
{\bf Nominal control law}. The nominal state feedback control law now is
\bea
u(t) \ts = \ts k_1^{*T} x(t) + k_2^* r_m(t) \nn \\
\ts = \ts  k_1^{*T} x(t) + k_{21}^{*T} \omega_{u_m}(t) +
k_{22}^{*T} \omega_{y_m}(t) + k_{20}^* y_m(t) + k_3^* u_m(t),
\label{5u1n1}
\eea
where 
\beq
k_{21}^{*} = k_2^* \beta_1,\;k_{22}^* = k_2^* \beta_2,\;k_{20}^* =
k_2^* \beta_{20},\;k_3^* = k_2^* \alpha_2.
\eeq

\medskip
{\bf Adaptive control law}. The adaptive state feedback control law is
\bea
u(t)\ts =\ts  k_1^{T} x(t) + k_{21}^{T} \omega_{u_m}(t) +
k_{22}^{T} \omega_{y_m}(t) + k_{20} y_m(t) + k_3 u_m(t)\nn\\
\ts = \ts \theta^T(t) \omega(t),
\label{5u1n1a}
\eea
where $k_1$, $k_{21}$, $k_{22}$, $k_{20}$ and $k_3$ are the estimates
of $k_1^*$, $k_{21}^{*}$, $k_{22}^*$, $k_{20}^*$ and $k_3^*$, and
\bea
\theta(t) \ts = \ts [k_1^T(t), k_{21}^T(t), k_{22}^T(t), k_{20}(t),
k_3(t)]^T \in R^{3n}\\
\omega(t) \ts = \ts [x^T(t), \omega_{u_m}^T(t), \omega_{y_m}^T(t),
y_m(t), u_m(t)]^T \in R^{3n}.
\eea

\medskip
{\bf Tracking error equation}. With the control law (\ref{5u1n1a}),
similar to (\ref{5e(t)}), the
closed-loop tracking error equation for $e(t) = y(t) - y_m(t)$ can be
derived as
\beq
e(t) = \rho^* W_m(s)[(\theta - \theta^*)^T \omega](t), 
\label{5e(t)1}
\eeq
which has the same form as that in (\ref{5e(t)}), with 
$
\theta^* = [k_1^{*T}, k_{21}^{*T}, k_{22}^{*T}, k_{20}^*,
k_3^*]^T \in R^{3n}.
$

\medskip
{\bf Adaptive parameter update laws}. 
Based on (\ref{5e(t)1}), we can define the estimation error as
\beq
\epsilon (t) =  e(t) + \rho(t) \xi(t),
\label{5epsilon20}
\eeq
where $\rho(t)$ is the estimate of $\rho^{*}$, and 
\bea
\xi(t) \ts = \ts \theta^{T}(t) \zeta(t) - W_m (s) [\theta^{T} \omega](t) \in R\\
\zeta (t) \ts = \ts W_{m}(s)[\omega](t)\in R^{3n},
\eea
and choose the adaptive laws for $\theta(t)$ and $\rho(t)$ as
\bea
\label{5thetadot2}
\dot{\theta} (t) \ts = \ts - \frac{\Gamma \sign[k_p] \zeta (t) \epsilon (t)} 
{m^2(t)} \\*[0.05in]
\dot{\rho} (t) \ts =  \ts- \frac{\gamma\, \xi(t) \epsilon (t)} 
{m^2(t)}
\label{5rhodot2}
\eea
where $\Gamma = \Gamma^T > 0$ and $\gamma > 0$, 
and $m(t) = \sqrt{1 + \zeta^{T}(t) \zeta(t) + \xi^{2}(t)}$. 

\medskip
This adaptive control scheme, with the completely parametrized
tracking error $e(t)$ in (\ref{5e(t)1}) and the linear form estimation
error $\epsilon(t)$: 
\beq
\epsilon (t) = \rho^{*} (\theta(t) - \theta^{*})^{T} \zeta(t) + (\rho(t) - 
\rho^{*}) \xi(t),
\label{5732A.362}
\eeq
also has the desired parameter adaptation properties, closed-system stability and
asymptotic output tracking: $\lim_{t \rightarrow \infty} (y(t) -
y_m(t)) = 0$, similar to that for the design with $x_m(t)$ in Section
\ref{Design Using Reference System State xm(t)}.

\setcounter{equation}{0}
\section{Output Feedback Control Designs}
\label{Output Feedback Control Designs}
Recall from (\ref{5732A.930}) and (\ref{5732A.951}) that the nominal
traditional output feedback controller structure is
\begin{equation}
u(t) = \theta_{1}^{*T} \omega_{1}(t) + \theta_{2}^{*T}
\omega_{2}(t) + \theta_{20}^* y(t) + \theta_{3}^* r(t),
\label{5732A.9311}
\end{equation}
where $\theta_{1}^* \in R^{n-1}$, $\theta_{2}^* \in R^{n-1}$, $\theta_{20}^*
\in R$ and $\theta_{3}^* \in R$, and 
\begin{equation}
\omega_{1}(t) = \frac{a(s)}{\Lambda(s)}[u](t),\;\omega_{2}(t) = 
\frac{a(s)}{\Lambda(s)}[y](t)
\end{equation}
with $a(s) = [1,s,\cdots,s^{n-2}]^{T}$ and $\Lambda(s)$ being a monic
stable polynomial of degree $n-1$. 

For the new adaptive control problem with an unknown reference system 
(\ref{5x0}): $\dot{x}_{m}(t) = A_{m} x_{m}(t) + b_{m} u_m(t),\;y_m(t) = c_m
x_m(t)$, the above signal $r(t)$ needs to be replaced by $r_m(t) =
P_m(D) G_m(D)[u_m](t) = P_m(D)[y_m](t)$ to fit in a MRAC scheme. Since
$r_m(t)$ is unavailable due to the uncertainties of $G_m(s) = c_m(sI -
A_m)^{-1} b_m$ or $y_m^{(i)}(t)$, $i=1,2,\ldots, n^*$, its
parametrization and estimation are also to be used, similar to a state
feedback control design.

\subsection{Design Using Reference System State $x_m(t)$}
As derived in (\ref{5r01}), in terms of $x_m(t)$ and $u_m(t)$, $r_m(t)$ can be parametrized as
\beq
r_m(t) = \alpha_1^T x_m(t) + \alpha_2 u_m(t)
\label{5rm(t)u}
\eeq
for some parameters $\alpha_1 \in R^n$ and $\alpha_2 \in R$ which depend 
on the unknown reference system parameters ($A_m$, $b_m$, $c_m$) and are 
to define the nominal controller parameters.

\bigskip
{\bf Nominal control law}. With $x_m(t)$ available, the nominal control law is
\bea
u(t) \ts = \ts \theta_{1}^{*T}\omega_{1}(t) + \theta_{2}^{*T}\omega_{2}(t) + \theta_{20}^*
 y(t) + \theta_{3}^* r_m(t) \nn\\
\ts = \ts \theta_{1}^{*T}\omega_{1}(t) + \theta_{2}^{*T}\omega_{2}(t) + \theta_{20}^*
 y(t) + \theta_{3}^* (\alpha_1^T x_m(t) + \alpha_2 u_m(t)) \nn\\
\ts = \ts \theta_{1}^{*T}\omega_{1}(t) + \theta_{2}^{*T}\omega_{2}(t) + \theta_{20}^*
 y(t) + \theta_{31}^{*T} x_m(t) + \theta_{32}^* u_m(t) \nn\\
\ts = \ts \theta^{*T}\omega(t), 
\label{5732A.931}
\eea
where 
\beq
\theta_{31}^* = \theta_3^* \alpha_1 \in R^n,\;
\theta_{32}^* = \theta_3^* \alpha_2 \in R
\eeq
\beq
\theta^* =
[\theta_1^{*T}, \theta_2^{*T}, \theta_{20}^*, \theta_{31}^{*T}, \theta_{32}^*]^T \in R^{3n}
\label{5thetasa}
\eeq
\beq
\omega(t) = [\omega_1^T(t), \theta_2^T(t), y(t), x_m^T(t),
u_m(t)]^T \in R^{3n}.
\label{5omegaa}
\eeq

Such a nominal control law can achieve the desired control objective:
all closed-loop system signals are bounded and
$\lim_{t \rightarrow \infty} (y(t) - y_m(t)) = 0$
(exponentially). When $\theta^*$ is unknown, adaptive
 control uses the estimates $\theta(t)$ of
$\theta^*$ to ensure $\lim_{t \rightarrow \infty} (y(t) - y_m(t)) =
0$ (asymptotically).

\bigskip
{\bf Adaptive control law and tracking error equation}. Based on the nominal control law 
(\ref{5732A.931}), the adaptive control law is constructed as
\beq
u(t) = \theta^T(t) \omega(t),
\label{5uu(t)1}
\eeq
where $\omega(t)$ is defined in (\ref{5omegaa}), and $\theta(t)$ is
the estimate of $\theta^*$ defined in (\ref{5thetasa}). 

\medskip
From (\ref{5732A.951}), we obtain
\bea
\ts \ts \theta_{1}^{*T} a(s) P(s)[y](t) 
+ (\theta_{2}^{*T} a(s) + \theta_{20}^{*} 
\Lambda(s)) k_{p} Z(s)[y](t)\nn\\
\ts  =\ts \Lambda(s)(P(s) - k_{p}  
\theta_{3}^{*} Z(s) P_{m}(s))[y](t),
\eea
and with $P(s)[y](t) = k_p Z(s)[u](t)$ as $y(t) = G(s)[u](t)$ for 
$G(s) = k_p \frac{Z(s)}{P(s)}$, and 
$F(s) = \frac{a(s)}{\Lambda(s)}$, we have 
\bea
\ts \ts \theta_{1}^{*T} F(s)[u](t) 
+ (\theta_{2}^{*T} F(s) + \theta_{20}^{*})[y](t) + \theta_{3}^{*} r(t)\nn\\
\ts  =\ts u(t) - \theta_{3}^{*} P_{m}(s)[y](t) + \theta_{3}^{*}
P_{m}(s)[y_m](t),
\label{matchinge}
\eea
where $P_{m}(s)[y_m](t) = r(t)$. Then, similar to (\ref{5tee0}), in
this case, with $\theta_{3}^{*} r(t) = \theta_{31}^{*T} x_m(t) +
\theta_{32}^* u_m(t)$ and (\ref{5uu(t)1}), the tracking error $e(t) =
y(t) - y_m(t)$ satisfies
\beq
e(t) = \rho^* W_m(s)[(\theta - \theta^*)^T \omega](t),
\label{5utee1}
\eeq
for $\rho^* = k_p$ and 
the different $\theta$, $\theta^*$ and $\omega(t)$, which has the same
form as (\ref{5e(t)}) and (\ref{5e(t)1}), and can be used to define the estimation error and
adaptive laws for $\theta(t)$ and the estimate $\rho(t)$ of $\rho^*$.

\bigskip
{\bf Adaptive parameter update laws}. Based on (\ref{5utee1}), we can define the estimation error
\beq
\epsilon (t) =  e(t) + \rho(t) \xi(t),
\label{5epsilon30}
\eeq
where $\rho(t)$ is the estimate of $\rho^{*}$, and 
\bea
\xi(t) \ts = \ts \theta^{T}(t) \zeta(t) - W_m (s) [\theta^{T} \omega](t) \in R\\
\zeta (t) \ts = \ts W_{m}(s)[\omega](t) \in R^{n_\theta}, 
\eea
and choose the adaptive laws for $\theta(t)$ and $\rho(t)$ as
\bea
\label{5uthetalaw1}
\dot{\theta}(t) \ts = \ts - \frac{\Gamma \sign[k_p] \zeta (t) \epsilon (t)} 
{m^2(t)} \\*[0.05in]
\dot{\rho}(t) \ts = \ts 
- \frac{\gamma\, \xi(t) \epsilon (t)}{m^2(t)},
\label{5urholaw1}
\eea
where $\Gamma = \Gamma^T > 0$, $\gamma > 0$, and
$
m(t) = \sqrt{1 + \zeta^{T}(t) \zeta(t) + \xi^{2}(t)}.
$

\medskip
This output feedback adaptive control scheme, with the completely parametrized
tracking error $e(t)$ in (\ref{5utee1}) and the linear estimation error
$\epsilon(t)$: 
\beq
\epsilon (t) = \rho^{*} (\theta(t) - \theta^{*})^{T} \zeta(t) + (\rho(t) - 
\rho^{*}) \xi(t),
\label{5732A.3621}
\eeq
ensured by the parametrization and estimation of the unknown $r_m(t)$
for the control laws (\ref{5732A.931}) and (\ref{5uu(t)1}), 
also has the desired parameter adaptation properties, closed-system stability and
asymptotic output tracking: $\lim_{t \rightarrow \infty} (y(t) -
y_m(t)) = 0$, similar to that for the state feedback adaptive control
design using the reference system state vector $x_m(t)$ in Section
\ref{Design Using Reference System State xm(t)}.

\subsection{Design Using Reference System Output $y_m(t)$}
\label{Design Using Reference System Output y_m(t)}
When $x_m(t)$ is not available, from (\ref{5r02}), we have
\beq
r_m(t) = \beta_1^T \omega_{u_m}(t) + \beta_{2}^T \omega_{y_m}(t) +
\beta_{20} y_m(t) + \alpha_2 u_m(t)
\label{5r02u}
\eeq
for some $\beta_1 \in R^{n-1}$, $\beta_{2} \in R^{n-1}$ and
$\beta_{20} \in R$, where
\beq
\omega_{u_m}(t) = \frac{a(s)}{\Lambda_e(s)}[u_m](t),\;\omega_{y_m}(t) = \frac{a(s)}{\Lambda_e(s)}[y_m](t),
\eeq
for $a(s) = [1, s, \ldots, s^{n - 2}]^{T}$ and a chosen monic stable
polynomial $\Lambda_e(s)$ of degree $n-1$.

\medskip
The parameters $\alpha_i$ and $\beta_j$ are used to define the
parameters of a nominal control law whose structure is crucial for
adaptive control for which adaptive parameter estimates are used.

\bigskip
{\bf Nominal control law}. We use (\ref{5r02u}) to construct the 
nominal control law
\bea
u(t) \ts = \ts \theta_{1}^{*T}\omega_{1}(t) + \theta_{2}^{*T}\omega_{2}(t) + \theta_{20}^*
 y(t) + \theta_{3}^* r_m(t) \nn\\
\ts = \ts \theta_{1}^{*T}\omega_{1}(t) + \theta_{2}^{*T}\omega_{2}(t) + \theta_{20}^*
 y(t) \nn\\
\ts \ts + \theta_{3}^* (\beta_1^T \omega_{u_m}(t) + \beta_{2}^T \omega_{y_m}(t) +
\beta_{20} y_m(t) + \alpha_2 u_m(t)) \nn\\
\ts = \ts \theta_{1}^{*T}\omega_{1}(t) + \theta_{2}^{*T}\omega_{2}(t) + \theta_{20}^*
 y(t) \nn\\
\ts \ts +  \theta_{31}^{*T} \omega_{u_m}(t) + \theta_{32}^{*T} \omega_{y_m}(t) +
\theta_{33}^* y_m(t) + \theta_{34}^* u_m(t)\nn\\
\ts = \ts \theta^{*T}\omega(t), 
\label{5732A.932}
\eea
where
\beq
\theta_{31}^* = \theta_{3}^* \beta_1,\;
\theta_{32}^* = \theta_{3}^* \beta_2,\;
\theta_{33}^* = \theta_{3}^* \beta_{20},\;
\theta_{34}^* = \theta_{3}^* \alpha_2
\label{5thetas3i} 
\eeq
\beq
\theta^* =
[\theta_{1}^{*T}, \theta_{2}^{*T}, \theta_{20}^*, \theta_{31}^{*T}, \theta_{32}^{*T}, \theta_{33}^*, \theta_{34}^*]^T \in R^{4n-1}
\label{5thetasu} 
\eeq
\beq
\omega(t) = [\omega_{1}^T(t), \omega_{2}^T(t), 
y(t), \omega_{u_m}^T(t), \omega_{y_m}^T(t), y_m(t), u_m(t)]^T.
\label{5omegau}
\eeq

\bigskip
{\bf Adaptive control law and tracking error equation}. 
Based on the nominal control law (\ref{5732A.932}), the adaptive
control law is chosen as
\beq
u(t) = \theta^T(t) \omega(t),
\label{5uu(t)2}
\eeq
where $\omega(t)$ is defined in (\ref{5omegau}), and $\theta(t)$ is
the estimate of $\theta^*$ defined in (\ref{5thetasu}). 

\medskip
From the matching equation (\ref{matchinge}), with $\theta_3^* r(t) = 
 \theta_{31}^{*T} \omega_{u_m}(t) + \theta_{32}^{*T} \omega_{y_m}(t) +
\theta_{33}^* y_m(t) + \theta_{34}^* u_m(t)$ and (\ref{5uu(t)2}), similar to
(\ref{5utee1}), in this case, the tracking error $e(t) = y(t) -
y_m(t)$ satisfies
\beq
e(t) = \rho^* W_m(s)[(\theta - \theta^*)^T \omega](t),
\label{5utee}
\eeq
for $\rho^* = k_p$ and the 
different $\theta$, $\theta^*$ and $\omega(t)$, which has the same
form as (\ref{5e(t)}), (\ref{5e(t)1}) and (\ref{5utee1}), and 
can be used to define the estimation error and
adaptive laws for $\theta(t)$ and the estimate $\rho(t)$ of $\rho^*$.

\bigskip
{\bf Adaptive parameter update laws}. 
Based on (\ref{5utee}), we also define the estimation error
\beq
\epsilon (t) =  e(t) + \rho(t) \xi(t),
\label{5epsilon4}
\eeq
where $\rho(t)$ is the estimate of $\rho^{*}$, and 
\bea
\xi(t) \ts = \ts \theta^{T}(t) \zeta(t) - W_m (s) [\theta^{T} \omega](t) \in R\\
\zeta (t) \ts = \ts W_{m}(s)[\omega](t) \in R^{n_\theta}. 
\eea
The estimation error $\epsilon(t)$ can be expressed as the linear error form
\beq
\epsilon (t) = \rho^{*} \tilde{\theta}^T(t) \zeta(t)
+ \tilde{\rho}(t) \xi(t),
\label{5lef}
\eeq
with $\tilde{\theta}(t) = \theta(t) - \theta^{*}$ and $\tilde{\rho}(t) = \rho(t) - \rho^{*}$. 

\medskip
We then choose the adaptive laws for $\theta(t)$ and $\rho(t)$ as
\bea
\label{5uthetalaw}
\dot{\theta}(t) \ts = \ts - \frac{\Gamma \sign[k_p] \zeta (t) \epsilon (t)} 
{m^2(t)} \\*[0.05in]
\dot{\rho}(t) \ts = \ts - \frac{\gamma\, \xi(t) \epsilon (t)}{m^2(t)},
\label{5urholaw}
\eea
where $\Gamma = \Gamma^T > 0$, $\gamma > 0$, and 
$
m(t) = \sqrt{1 + \zeta^{T}(t) \zeta(t) + \xi^{2}(t)}.
$

\medskip
This output feedback adaptive control scheme using the reference
system output $y_m(t)$ has the similar parameter adaptation properties 
(to be shown in Lemma \ref{5lemma11}) and closed-loop stability and
asymptotic output tracking properties (to be shown in Theorem \ref{5thm1}).

\setcounter{equation}{0}
\section{Adaptive Control System Properties}
Thus far, we have developed four adaptive control schemes for the adaptive output 
tracking control problem with the uncertain $G_m(s) = c_m (sI - A_m)^{-1} b_m$ 
and $y_m^{(i)}(t)$, $i=1,2,\ldots, n^*$, of the reference model system 
(\ref{5x0}): $\dot{x}_{m}(t) = A_{m} x_{m}(t) + b_{m} u_m(t),\;y_m(t) = c_m
x_m(t)$, that is, with the uncertain $r_m(t) = P_m(D) G_m(D)[u_m](t)
 = P_m(D)[y_m](t)$ which has been a key component of a traditional model 
reference adaptive controller, but not available for the current problem.

\medskip
We have proposed a parametrization and estimation solution to deal with 
such an uncertain $r_m(t)$, for developing the four adaptive control schemes: 
(1) state feedback control design using $x_m(t)$, 
(2) state feedback control design using $y_m(t)$, 
(3) output feedback control design using $x_m(t)$, and 
(4) output feedback control design using $y_m(t)$, with all schemes using $u_m(t)$.

\medskip
{\bf Unified form of adaptive control schemes}. 
The four adaptive control systems have the unified control law form: 
$u(t) = \theta^T \omega(t)$ (see (\ref{5u1a}), (\ref{5u1n1a}),
(\ref{5uu(t)1}), (\ref{5uu(t)2}), similar to the traditional ones
(\ref{5732A.930}) and (\ref{5sfcs})), 
as well as the tracking error equation form: $e(t) = y(t) -
y_m(t) = \rho^* W_m(s)[(\theta - \theta^*)^T \omega](t)$ (see (\ref{5e(t)}),
(\ref{5e(t)1}), (\ref{5utee1}), (\ref{5utee}), similar to the
traditional ones (\ref{5tee0}) and (\ref{5tee00})), the estimation error
form: $\epsilon(t) = e(t) + \rho(t) \xi(t)$ (see (\ref{5732A.35}),
(\ref{5epsilon20}), (\ref{5epsilon30}), (\ref{5epsilon4})) and its linear error form: 
$\epsilon (t) = \rho^{*} (\theta(t) - \theta^{*})^{T} \zeta(t) +
(\rho(t) - \rho^{*}) \xi(t)$ (see (\ref{5732A.36}), (\ref{5732A.362}),
(\ref{5732A.3621}), (\ref{5lef})), 
and the adaptive parameter update laws (see (\ref{5732A.37})-(\ref{5732A.38}), 
(\ref{5thetadot2})-(\ref{5rhodot2}),
(\ref{5uthetalaw1})-(\ref{5urholaw1}),
(\ref{5uthetalaw})-(\ref{5urholaw})), which all have a unified form
for all four adaptive control schemes.

\subsection{Adaptive Law Properties}
Based on their unified forms, we can analyze the four sets of adaptive parameter update laws by 
deriving the time-derivative of the positive definite function
$
V(\tilde{\theta}, \tilde{\rho}) = |\rho^{*}|\tilde{\theta}^{T} 
\Gamma^{-1} \tilde{\theta} + \gamma^{-1} \tilde{\rho}^{2}
$ as
\beq
\dot{V} = - \frac{2 \epsilon^2(t)}{m^2(t)} \leq 0,
\label{5Vdot}
\eeq
based on which we can derive the following desired properties.

\begin{lem} 
\label{5lemma11}
The four sets of adaptive parameter update laws for the four designed
adaptive control schemes ensure that 
$\theta(t) \in L^{\infty}$, $\rho(t) \in L^{\infty}$, 
 $\frac{\epsilon(t)}{m(t)} \in L^{2} \cap L^{\infty}$,
 $\dot{\theta}(t) \in L^{2} \cap L^{\infty}$, and 
 $\dot{\rho}(t) \in L^{2} \cap L^{\infty}$.
\end{lem} 

The properties $\theta(t) \in L^{\infty}$, $\rho(t) \in L^{\infty}$
and $\frac{\epsilon(t)}{m(t)} \in L^{2}$ follow directly from
(\ref{5Vdot}), the property $\frac{\epsilon(t)}{m(t)} \in L^{\infty}$
follows from $\epsilon (t) = \rho^{*} (\theta(t) - \theta^{*})^{T} \zeta(t) +
(\rho(t) - \rho^{*}) \xi(t)$ and $m(t) = \sqrt{1 + \zeta^{T}(t)
  \zeta(t) + \xi^{2}(t)}$, and $\dot{\theta}(t) \in L^{2} \cap L^{\infty}$ and 
 $\dot{\rho}(t) \in L^{2} \cap L^{\infty}$ follow from the established
properties.

\subsection{Closed-loop System Properties}
We can then use the obtained $L^2$ and $L^\infty$ properties of the
adaptive parameter update laws, to establish the following desired
stability and asymptotic tracking properties for the adaptive control
systems with the four corresponding adaptive control schemes in the four
cases studied in Section 2 and Section 3: state feedback control using
$x_m(t)$, state feedback control using $y_m(t)$, output feedback
control using $x_m(t)$, and output feedback control using $y_m(t)$,
for the current model reference adaptive control (MRAC) problem with
reference system uncertainties.

\begin{thm}
\label{5thm1}
The four adaptive controllers, updated from their adaptive laws and
applied to the plant (\ref{5x1}), ensure that all closed-loop
system signals are bounded and $\lim_{t \rightarrow \infty} (y(t) - y_m(t)) = 0$.
\end{thm}
{\bf Proof}: The proof consists of three parts: the stability and
asymptotic output tracking results for a traditional output feedback
MRAC system, the connection of the current output feedback adaptive
control system to the traditional MRAC system, and the connection of
the current state feedback adaptive control system to the current
output feedback adaptive control system.

\bigskip
{\bf Traditional MRAC}. The closed-loop system stability and asymptotic output
tracking of a traditional MRAC scheme is proved in \cite[pages
  216-218]{t03}, with the controller structure (\ref{5732A.930}):
\begin{equation}
u(t) = \theta_{1}^{T}\omega_{1}(t) + \theta_{2}^{T}\omega_{2}(t) + \theta_{20}
 y(t) + \theta_{3} r_m(t),
\label{5732A.93a}
\end{equation}
as in (\ref{5732A.930}),
 where $\theta_{1} \in R^{n-1}$, $\theta_{2} \in R^{n-1}$, $\theta_{20}
\in R$ and $\theta_{3} \in R$, and 
\begin{equation}
\omega_{1}(t) = \frac{a(s)}{\Lambda(s)}[u](t),\;\omega_{2}(t) = 
\frac{a(s)}{\Lambda(s)}[y](t)
\end{equation}
with $a(s) = [1,s,\cdots,s^{n-2}]^{T}$ and $\Lambda(s)$ being a monic
stable polynomial of degree $n-1$. 

Two key features of the proof are:
the $L^\infty$ properties of the adaptive laws ensure a feedback
framework of the closed-loop system, and the $L^2$ properties of the
adaptive laws ensure that the loop gain of the feedback framework is
small, so that the closed-loop system is stable, that is, all signals
are bounded. The $L^2$ properties then also imply that $\lim_{t
  \rightarrow \infty} (y(t) - y_m(t)) = 0$.

\medskip
\medskip
{\bf Current output feedback MRAC designs}. In the current MRAC problem, $r_m(t) = P_m(s) G_m(s)[u_m](t)$
is unknown as $G_m(s) = c_m(s I - A_m)^{-1} b_m$ is unknown. We 
have parametrized it either, for $x_m(t)$ available, as 
$r_m(t) = \alpha_1^T x_m(t) + \alpha_2 u_m(t)$ as in (\ref{5rm(t)u}),
or, for $x_m(t)$ unavailable, as $r_m(t) = \alpha_1^T \hat{x}_m(t) +
\alpha_2 u_m(t)$, for an observer-based estimate
$\hat{x}_m(t)$ of $x_m(t)$ (see (\ref{5hatxm})), to obtain
\bea
r_m(t) \ts = \ts \alpha_1^T \hat{x}_m(t) + \alpha_2 u_m(t) \nn\\
\ts = \ts \beta_1^T \omega_{u_m}(t) + \beta_{2}^T \omega_{y_m}(t) +
\beta_{20} y_m(t) + \alpha_2 u_m(t),
\eea
as in (\ref{5r02u}), for some $\beta_1 \in R^{n-1}$, $\beta_{2} \in R^{n-1}$ and
$\beta_{20} \in R$, where
\beq
\omega_{u_m}(t) = \frac{[1, s, \ldots, s^{n - 2}]^{T}}{\Lambda_e(s)}[u_m](t),\;\omega_{y_m}(t) = \frac{[1, s, \ldots, s^{n - 2}]^{T}}{\Lambda_e(s)}[y_m](t),
\eeq
with a chosen monic stable polynomial $\Lambda_e(s)$ of degree $n-1$.

From (\ref{5732A.931}), the adaptive controller structure for $x_m(t)$
available is
\beq
u(t) =  \theta_{1}^{T}\omega_{1}(t) + \theta_{2}^{T}\omega_{2}(t) + \theta_{20}
 y(t) + \theta_{31}^{T} x_m(t) + \theta_{32} u_m(t), 
\label{5732A.931a}
\eeq 
and from (\ref{5732A.932}), the adaptive controller structure for $x_m(t)$
unavailable is
\bea
u(t) \ts = \ts \theta_{1}^{T}\omega_{1}(t) + \theta_{2}^{T}\omega_{2}(t) + \theta_{20}
 y(t) \nn\\
\ts \ts +  \theta_{31}^{T} \omega_{u_m}(t) + \theta_{32}^{T} \omega_{y_m}(t) +
\theta_{33} y_m(t) + \theta_{34} u_m(t).
\label{5732A.932a}
\eea
Comparing (\ref{5732A.931a}) and (\ref{5732A.932a}) with
(\ref{5732A.93a}), we can see that they all have the common parts:  
$\theta_{1}^{T}\omega_{1}(t) + \theta_{2}^{T}\omega_{2}(t)
+ \theta_{20} y(t)$ which depend on $u(t)$ and $y(t)$ and are crucial for system stability, while
their other parts: $\theta_{31}^{T} x_m(t) + \theta_{32} u_m(t)$, 
$\theta_{31}^{T} \omega_{u_m}(t) + \theta_{32}^{T} \omega_{y_m}(t) +
\theta_{33} y_m(t) + \theta_{34} u_m(t)$ and $\theta_3 r_m(t)$ are all
bounded and do not affect the system stability (signal boundedness).
Hence, under the established $L^2$ properties, 
the analysis procedure of \cite[pages 216-218]{t03} is
applicable to the adaptive controller (\ref{5732A.931a}) for $x_m(t)$
available and (\ref{5732A.932a}) for $x_m(t)$ unavailable, to conclude
the results of Theorem \ref{5thm1}.

\bigskip
{\bf Current state feedback MRAC designs}. From (\ref{5u1a}),
the adaptive state feedback control law with $x_m(t)$ available is
\bea
u(t) \ts = \ts  k_1^{T} x(t) + 
k_{21}^{T} x_m(t) + k_{22} u_m(t),
\label{5u1a1}
\eea
and, from (\ref{5u1n1a}), the adaptive state feedback control law with
$x_m(t)$ unavailable is
\bea
u(t)\ts =\ts  k_1^{T} x(t) + k_{21}^{T} \omega_{u_m}(t) +
k_{22}^{T} \omega_{y_m}(t) + k_{20} y_m(t) + k_3 u_m(t).
\label{5u1n1a1}
\eea

To use the stability analysis framework established in 
\cite{t03} (as described above for the traditional MRAC system), we
express $x(t)$ in terms of $u(t)$ and $y(t)$, using a formulation
based on a fictitious estimate $\hat{x}(t)$ of $x(t)$, designed using
the reduced-order observer theory \cite{r96} as
\beq
\hat{x}(t) = Q \left[\begin{array}{c}
y(t)\\
w(t) + L_r y(t)
\end{array}
\right],
\eeq
for the plant: 
$\dot{x}(t) = A x(t) + b\, u(t), y(t) = c\, x(t)$,
such that $\lim_{t \rightarrow \infty} (\hat{x}(t) - x(t)) = 0$ exponentially, 
where $Q \in R^{n\times n}$ is a nonsingular matrix such that $y(t)
= [1, 0, \ldots, 0] Q^{-1} x(t)$, $L_r \in R^{n-1}$ is a reduced-order observer gain vector, and the
dynamic part $w(t)$ of $\hat{x}(t)$ is
\beq
w(t) = \frac{G_{u}(s)}{\Lambda(s)}[u](t) + 
 \frac{G_{y}(s)}{\Lambda(s)}[y](t) \in R^{n-1}
\eeq
for a chosen monic stable polynomial $\Lambda(s)$ of degree
$n-1$, and some $(n-1)$-dimensional polynomial vectors $G_{u}(s)$
and $G_{y}(s)$.

\medskip
With $k_1(t)$ bounded, we have 
\beq
\lim_{t \rightarrow \infty} (k_1^T(t) \hat{x}(t) - k_1^T(t) x(t)) = 0,
\eeq
exponentially, so that the control law (\ref{5u1a1}) is equivalent to
the observer-based control law:
\bea
u(t) \ts = \ts  k_1^{T} \hat{x}(t) + 
k_{21}^{T} x_m(t) + k_{22} u_m(t) \nn\\
\ts = \ts \theta_1^{T} \omega_1(t) + \theta_2^T \omega_2(t)
+ \theta_{20} y(t) + k_{21}^{T} x_m(t) + k_{22} u_m(t),
\label{5u1a2}
\eea
for some $\theta_{1} \in R^{n-1}$, $\theta_{2} \in R^{n-1}$ and $\theta_{20}
\in R$, and, with $a(s) = [1,s,\cdots,s^{n-2}]^{T}$, 
\begin{equation}
\omega_{1}(t) = \frac{a(s)}{\Lambda(s)}[u](t),\;\omega_{2}(t) = 
\frac{a(s)}{\Lambda(s)}[y](t). 
\end{equation}
We can see that (\ref{5u1a2}), with bounded 
$k_{21}^{T} x_m(t) + k_{22} u_m(t)$, has the same structure as that in
(\ref{5732A.931a}) with bounded $\theta_{31}^{T} x_m(t) + \theta_{32} u_m(t)$.

\medskip
The same idea leads the control law (\ref{5u1n1a1}) to its equivalent version:
\bea
u(t)\ts =\ts  \theta_1^{T} \omega_1(t) + \theta_2^T \omega_2(t)
+ \theta_{20} y(t) \nn\\
\ts \ts + k_{21}^{T} \omega_{u_m}(t) +
k_{22}^{T} \omega_{y_m}(t) + k_{20} y_m(t) + k_3 u_m(t),
\label{5u1n1a2}
\eea
which, with $k_{21}^{T} \omega_{u_m}(t) +
k_{22}^{T} \omega_{y_m}(t) + k_{20} y_m(t) + k_3 u_m(t)$ bounded, has
the same structure as that in (\ref{5732A.932a}) with $\theta_{31}^{T}
\omega_{u_m}(t) + \theta_{32}^{T} \omega_{y_m}(t) + \theta_{33} y_m(t)
+ \theta_{34} u_m(t)$ bounded.

\medskip
Moreover, for both adaptive control laws (\ref{5u1a2}) and (\ref{5u1n1a2}), the total parameter vector
\beq
\theta(t) = [\theta_1^T(t), \theta_2^T(t), \theta_{20}(t), \ldots]^T
\eeq
has the desired properties: $\theta(t) \in L^\infty$ and
$\dot{\theta}(t) \in L^2 \cap L^\infty$, as the other components already have. 

\medskip
Then, with the additional $L^\infty$ and $L^2$ properties of the
adaptive laws, the analysis procedure of \cite[pages 216-218]{t03} can
also be applied to (\ref{5u1a2}) with $x_m(t)$ or (\ref{5u1n1a2}) for
$x_m(t)$ unavailable, to conclude the results of Theorem \ref{5thm1},
for the state feedback control laws (\ref{5u1a1}) and
 (\ref{5u1n1a1}). \hspace*{\fill} $\nabla$ 

\begin{rem}
\rm
For the reference model system (\ref{5x0}): 
$\dot{x}_{m}(t) = A_{m} x_{m}(t) + b_{m} u_m(t),\;y_m(t) = c_m
x_m(t)$, or $y_m(t) = G_m(s)[u_m](t)$, $G_m(s)$ does not need to be stable as
long as $u_m(t)$ stabilizes the reference model system to ensure that $y_m(t)$,
$x_m(t)$ and $u_m(t)$ are bounded. 

For example, we use an output feedback control law: 
$u_m(t) = H_m(s)[y_m](t) + v_m(t)$ with $v_m(t)$ bounded and $H_m(s)$
stabilizing, such that $y_m(t) = G_m(s)(1 - H_m(s)
G_m(s))^{-1}[v_m](t)$, where $G_m(s)(1 - H_m(s) G_m(s))^{-1}$ is
stable, so that $r_m(t) = P_m(s) G_m(s)(1 - H_m(s)
G_m(s))^{-1}[v_m](t)$ is bounded, as $P_m(s) G_m(s)(1 - H_m(s)
G_m(s))^{-1}$ is proper or strictly proper, given that $P_m(s)
G_m(s)$ is proper or strictly proper. Similarly, we can use a state feedback
 stabilizing control law $u_m(t) = k_{1m}^T x_m(t) + v_m(t)$ for a gain
vector $k_{1m} \in R^n$. While the reference system $(A_m,
b_m, c_m)$ is unknown to the plant (\ref{5x1}) and its adaptive
control design, the knowledge of $(A_m, b_m, c_m)$ may be considered
to be available for the design of the control law $u_m(t)$ stabilizing
the reference system. 

Moreover, the stabilizing control law $u_m(t)$ may be from an adaptive
control law, designed with the knowledge of the relative degree
$n_m^*$ of $G_m(s)$ for the reference system under the standard MRAC
conditions (see Assumption (A1) in Section \ref{Technical
  Background}), without using the knowledge of $(A_m,
b_m, c_m)$ (or $G_m(s)$) and $y_r^{(i)}(t)$, $i=1,2,\ldots, n^*$, for
 $y_r(t)$ satisfying: $y_r(s) = W_r(s)[r](t)$ (as the reference model system
for the original reference system (\ref{5x0})), similar to
(\ref{5mrs}) (as the reference model system for the plant (\ref{5x1})), but
with $r(t)$ being a chosen signal.

The control signal $u_m(t)$ may also be from a pilot
 stabilizing the reference system (\ref{5x0}).
\hspace*{\fill} $\Box$ 
\end{rem}

\begin{rem}
\rm
If the reference model system is an unknown nonlinear system:
$\dot{x}_m(t) = f_m(x_m(t)) + g_m(x_m(t)) u_m(t),\;y_m(t) =
h_m(x_m(t))$, such as a feedback linearizable nonlinear system
\cite{sb89}, the similar ideas can be used to express 
$y_m^{(i)}(t)$, $i=1,2,\ldots, n^*$, in terms of $x_m(t)$ and
$u_m(t)$, to form a parametrization and estimation expression of
$r_m(t) = P_m(s)[y_m](t)$ for a chosen stable polynomial $P_m(s)$ of
degree $n^*$, to be included as the estimate of $r_m(t)$ in the
adaptive control law. 
\hspace*{\fill} $\Box$ 
\end{rem}

\subsection{Extensions}
The main technical novelty of this work is that the newly developed
adaptive control technique of parametrizing and estimating the
reference system uncertainties is able to ensure the asymptotic output
tracking a reference signal generated from a reference model system 
uncertain in either its output derivatives or dynamics which have been the
key information used in the existing model reference adaptive control
designs. The new adaptive control design technique, using an expanded
adaptive controller structure to deal with uncertain reference
systems, can be used to expand the tracking capacity of various
adaptive control systems in which the reference signals are the
outputs of some given reference systems with similar uncertainties.

\subsection{Partial-State Feedback Control Designs}
For the system (\ref{5x1}): $\dot{x}(t) = A x(t) + b u(t),\;y(t) = c
x(t)$, a partial-state feedback control scheme uses a partial state vector
$y_0(t) = C_0 x(t) \in R^{n_0}$, in stead of $x(t)$ or $y(t)$, for the
control signal design, where $(A,C_0)$ is observable with $C_0 \in
R^{n_0 \times n}$ and $\rank[C_0] = n_0$.

\medskip
\medskip
{\bf Partial-state feedback adaptive control}. 
As shown in \cite{st20}, with the available signal $y_0(t) = C_0 x(t)$, 
a partial-state (reduced-order) observer based estimate
$\hat{x}(t)$ of $x(t)$ can be designed such that 
$\lim_{t \rightarrow \infty} (\hat{x}(t) - x(t)) = 0$ exponentially,
so that the state feedback control law $u(t) = k_1^{*T} x(t) +
k_2^*r(t)$ can be replaced by the observer-based nominal control law
$u(t) = k_1^{*T} \hat{x}(t) + k_2^*r(t)$ which, based on the observer
construction, can be expressed as
\bea
u(t) \ts  = \ts \theta^{*T}_1\frac{a_1(s)}{\Lambda(s)}[u](t) +
\theta_2^{*T}\frac{A_2(s)}{\Lambda(s)}[y_0](t) + \theta_{20}^{*T}y_0(t) \nn\\
\ts \ts + \theta_3^*r(t) + \varepsilon_0(t), 
\eea
for a chosen monic stable polynomial $\Lambda(s)$ of degree $n-n_0$, 
and some exponentially decaying and initial condition related signal
$\varepsilon_0(t)$, parameter vectors 
$\theta_{1}^* \in R^{n-n_0}$, $\theta_2^* \in R^{n_0(n-n_0)}$,
$\theta_{20}^* \in R^{n_0}$ and $\theta_3^* \in R$, and 
$a_1(s) = [1, s,\ldots,s^{n-n_0-1}]^{T}, \,
A_2(s) = [I_{n_0},sI_{n_0},\ldots,s^{n-{n_0}-1}I_{n_0}]^{T}$.

Ignoring the decaying term $\varepsilon_0(t)$, we choose the adaptive
partial-state feedback controller structure:
\begin{equation}
u(t) = \theta_1^{*T}\omega_1(t) + \theta_2^{*T}\omega_2(t) +
\theta_{20}^{*T}y_0(t) + \theta_3^*r(t),
\label{psfnc}
\end{equation}
where $\omega_1(t) = \frac{a_1(s)}{\Lambda(s)}[u](t),\,\omega_2(t) =
\frac{A_2(s)}{\Lambda(s)}[y_0](t)$, and 
$\theta_1$, $\theta_2$, $\theta_{20}$ and $\theta_3$ are the estimates
of $\theta_1^{*}$, $\theta_2^{*}$, $\theta_{20}^{*}$ and
$\theta_3^*$. As shown in \cite{st20}, an adaptive law can be designed
to update the parameter estimates to ensure the closed-loop signal
boundedness and asymptotic output tracking of $y_m(t) = W_m(s)[r](t)$
for $W_m(s) = \frac{1}{P_m(s)}$ with $P_m(s)$ being a known stable
polynomial of degree $n^*$.

In this case, we have the matching equation 
\beq
G(s)\big(1-\theta_1^{*T}\frac{a_1(s)}{\Lambda(s)} -
(\theta_2^{*T}\frac{A_2(s)}{\Lambda(s)} + \theta_{20}^{*T})G_0(s)\big)^{-1}\theta_3^* = 
W_m(s),
\eeq
where $G(s) = k_p\frac{Z(s)}{P(s)}$ for $y(t) = G(s)[u](t)$ and
$G_0(s) = \frac{Z_0(s)}{P(s)}$ for $y_0(t) = G_0(s)[u](t)$, or the
matching equation
\beq
1-\theta_1^{*T}\frac{a_1(s)}{\Lambda(s)} -
(\theta_2^{*T}\frac{A_2(s)}{\Lambda(s)} + \theta_{20}^{*T})G_0(s) =
\theta_3^*W_m^{-1}(s)G(s),
\label{matchinge2}
\eeq
which is equivalent to, as compared to (\ref{5732A.951}), the matching equation
\bea
\ts \ts \theta_1^{*T}a_1(s)P(s) + (\theta_2^{*T}A_2(s) +
\theta_{20}^{*T}\Lambda(s))Z_0(s) \nn\\
\ts = \ts \Lambda(s)(P(s) -
k_p\theta_3^*Z(s)P_m(s)).
\eea

From the matching equation (\ref{matchinge2}), we obtain the matching
identity:
\beq
u(t)-\theta_1^{*T}\frac{a_1(s)}{\Lambda(s)}[u](t) -
(\theta_2^{*T}\frac{A_2(s)}{\Lambda(s)} + \theta_{20}^{*T})[y_0](t) =
\theta_3^*W_m^{-1}(s)[y](t).
\label{matchinge3}
\eeq

The adaptive version of the control law (\ref{psfnc}) is
\begin{equation}
u(t) = \theta_1^{T}\omega_1(t) + \theta_2^{T}\omega_2(t) +
\theta_{20}^{T}y_0(t) + \theta_3 r(t),
\label{psfnca}
\end{equation}
where $\theta_1$, $\theta_2$, $\theta_{20}$ and $\theta_3$ are the
adaptive estimates of $\theta_1^*$, $\theta_2^*$, $\theta_{20}^*$ and
$\theta_3^*$ respectively. With this adaptive control law, from
(\ref{matchinge3}), we can obtain the tracking error equation
(\ref{5tee0}) with $y(t)$ being replaced by $y_0(t)$, and develop a
stable adaptive control scheme. 

Since $y_0(t) = C_0 x(t) \in R^{n_0}$ can be either a vector
containing $y(t) = C x(t)$ or not, or a scalar $y_0(t) \not = y(t)$, a
partial-state feedback adaptive controller may provide certain extra
design flexibility, as compared with the state feedback controller
(\ref{5sfcs}). Since $n_0 < n$, it may also reduce control system
complexability (the number of integrators needed to implement the
adaptive controller), as compared with the output feedback controller
(\ref{5732A.930}).

\medskip
\medskip
{\bf Designs for uncertain reference model systems}. 
For the current adaptive control problem, the reference
output $y_m(t)$ is generated from a given reference system
(\ref{5x0}): $\dot{x}_{m}(t) = A_{m} x_{m}(t) + b_{m} u_m(t),\;y_m(t)
= c_m x_m(t)$, with the system matrices $A_{m} \in R^{n \times n}$,
$b_{m} \in R^{n}$ and $c_m \in R^{1 \times n}$ unknown or the
time-derivatives $y_m^{(i)}(t)$, $i=1,2,\ldots, n^*$, unavailable,
leading to an unavailable $r(t) = r_m(t)$ in (\ref{psfnc}) whose term $\theta_3^*
r_m(t)$ needs to be expanded, parametrized and estimated.

\medskip
For the partial-state feedback control law (\ref{psfnc}) with $r(t) =
r_m(t)$, there are three ways to expand the term $\theta_3^* r_m(t) =
\theta^* P_m(s) [y_m](t)$: 

\medskip
(i) For a design using $x_m(t)$, as in (\ref{5r01}), we use
\beq
r_m(t) = \alpha_1^T x_m(t) + \alpha_2 u_m(t).
\label{rm1}
\eeq

\medskip
(ii) For a design using $y_m(t)$, as in (\ref{5r02}), we use
\beq
r_m(t) = \beta_1^T \omega_{u_m}(t) + \beta_{2}^T \omega_{y_m}(t) +
\beta_{20} y_m(t) + \alpha_2 u_m(t).
\eeq

\medskip
(iii) For a design using a partial-state signal $y_{m0}(t) = C_{m0}
x_m(t) \in R^{n_m}$ with $n_m < n$, as similar to 
$\theta_1^{*T}\omega_1(t) + \theta_2^{*T}\omega_2(t) +
\theta_{20}^{*T}y_0(t)$ in (\ref{psfnc}), we use
\beq
r_m(t) = \gamma_1^T \omega_{u_m}(t) + \gamma_{2}^T \omega_{y_{m0}}(t) +
\gamma_{20}^T y_{m0}(t) + \alpha_2 u_m(t),
\eeq
for some parameter vectors $\gamma_1$, $\gamma_2$ and $\gamma_3$,
derived using $r_m(t) = \alpha_1^T \hat{x}_m(t) + \alpha_2 u_m(t)$,
where $\hat{x}_m(t)$ is the estimate of $x_m(t)$ in (\ref{rm1}), from a state
observer designed with the partial-state signal $y_{m0}(t)$.

\medskip
For all three expansions of $r_m(t)$, an adaptive control law, expanded from
(\ref{psfnca}), can be constructed, with which a tracking error
equation for $e(t) = y(t) - y_m(t)$, similar to (\ref{5utee}), can be
derived, and an estimation error $\epsilon(t)$ can then be introduced
for the design of a stable adaptive law to achieve the desired control
objective.

\subsection{Other Extensions}
We now address the extensions of the developed new adaptive control
schemes to some other systems.

\bigskip
{\bf Extensions to multi-input multi-output (MIMO) systems}. For a
MIMO system (plant): 
\beq
\dot{x}(t) = A x(t) + B u(t),\;y(t) = C x(t),\;t \geq 0
\label{42.2m}
\eeq
with $x(t) \in R^n$, $u(t) \in R^M$ and $y(t) \in R^M$, and 
constant $A \in R^{n\times n}$, $B \in R^{n \times M}$ and $C \in R^{M
  \times n}$, its transfer matrix is $G(s) = C(sI - A)^{-1} B$ which
has a modified interactor matrix $\xi_m(s)$ (a polynomial matrix) such
that $\lim_{s \rightarrow \infty} \xi_m(s) G(s) = K_p \in R^{M \times M}$ is
finite and nonsingular \cite{t03}, and $\xi_m^{-1}(s)$ is a stable
transfer matrix. For traditional MRAC, a reference model system
equation is introduced as
\beq
\xi_m(s)[y_m](t) = r(t),
\label{me1}
\eeq
where $y_m(t)$ is the reference output for the plant output $y(t)$ to
track, and $r(t)$ is a reference input which is also to be used in
the control law, indicating $\xi_m(s)[y_m](t)$ needs to be known.

\medskip
However, if the reference system is given as
\beq
\dot{x}_m(t) = A_{m} x_{m}(t) + B_{m} u_m(t),\;y_m(t) = C_m x_m(t), 
\label{42.3m}
\eeq
with known $x_m(t) \in R^n$, $u_m(t) \in R^M$ and $y_M(t) \in R^M$, and 
unknown constant matrices $A_M \in R^{n\times n}$, $B_M \in R^{n
  \times M}$ and $C_M \in R^{M \times n}$, the standard reference
system model (\ref{me1}) may not be satisfied. The new adaptive
control designs developed in Section \ref{State Feedback Control
  Designs} and Section \ref{Output Feedback Control Designs} can be
extended to MIMO systems, using multivariable adaptive control
techniques \cite{t03}, as shown in \cite{t24}, 
if $\lim_{s \rightarrow \infty} \xi_m(s) G_m(s)
= K_m \in R^{M \times M}$ is finite (not necessarily nonsingular), for
$G_m(s) = C_m(sI - A_m)^{-1} B_m$. A partial-state feedback control
design can also be developed, using the theory in \cite{st21}.

\bigskip
{\bf Extensions to discrete-time systems}. Similar results can be
obtained for model reference adaptive control of a discrete-time
LTI system (plant) with $y(t) \in R^M$ and $u(t) \in
R^M$:
\beq
x(t+1) = A x(t) + B u(t),\;y(t) = C x(t),\;
t = 0, 1, 2, \ldots,
\label{42.2md}
\eeq
for which, the plant output $y(t)$ is to track the output $y_m(t)$
of a given reference system
\beq
x_m(t+1) = A_{m} x_{m}(t) + B_{m} u_m(t),\;y_m(t) = C_m x_m(t),
\label{42.3md}
\eeq
with known $x_m(t) \in R^n$, $u_m(t) \in R^M$ and $y_M(t) \in R^M$,
and, unlike the traditional adaptive control problem, with 
unknown $A_M \in R^{n\times n}$, $B_M \in R^{n
  \times M}$ and $C_M \in R^{M \times n}$, or unknown $y_m(t+1)$ etc.

\medskip
For the discrete-time transfer matrix is $G(z) = C(zI - A)^{-1} B$
case, the modified interactor matrix $\xi_m(z)$ can be defined in a similar
way: $\lim_{z \rightarrow \infty} \xi_m(z) G(z) = K_p \in R^{M \times M}$ is
finite and nonsingular \cite{gs84}, \cite{t03}, with $\xi_m^{-1}(z)$
stable. The new model reference adaptive control results, developed
above for continuous-time systems, can be extended to discrete-time
systems, as shown in \cite{t24}, if $\lim_{z \rightarrow \infty} \xi_m(z) G_m(z)
= K_m \in R^{M \times M}$ is finite, for $G_m(z) = C_m(zI - A_m)^{-1}
B_m$. To deal with the uncertainty of $G_m(z)$ or $r(t) =
\xi_m(z)[y_m](t)$ which depends on some future values of $y_m(t)$, the
adaptive controller structure can be modified with parametrized
estimation of $r(t)$, to ensure a completely parametrized tracking
error equation for designing a stable adaptive controller.

\medskip
In both the continuous-time and discrete-time
cases, the reference output signal $y_m(t)$ can also be generated from
a nonlinear system with unknown parameters, and the unknown
derivatives (or future values) of $y_m(t)$ can be estimated in a
modified adaptive controller.

\bigskip
{\bf Extensions to nonlinear systems}. The results can also be
extended to adaptive tracking control of feedback linearizable
nonlinear systems and parametric-strict-feedback (or output-feedback)
nonlinear systems, with unknown derivatives of the reference output
signal from a given reference system with uncertain parameters. While
the information of such derivative signals is crucial for traditional
adaptive control designs, modified adaptive feedback linearization and
backstepping control designs, with parametrized estimation of the
unknown reference output derivatives, can deal with such uncertainty
of the reference output signal which can be from either a linear
reference system or a nonlinear reference system \cite{t24}. 

The adaptive tracking control problem can also be solved for robot
following control, where the leader robot acceleration signal
$\ddot{y}_m(t)$ and system parameters are unknown to the follower
robot, as studied in \cite{t24}, while such a $\ddot{y}_m(t)$ is
needed for the traditional adaptive robot control design.

\setcounter{equation}{0}
\section{Simulation Study}
In this section, we present a simulation study in which the reference
system is a leader aircraft and the plant is a follower aircraft, both
have the same dynamic model. The leader aircraft is controlled by a
built-in state feedback control law and the follower aircraft is
controlled by an output feedback adaptive control law developed in
Section \ref{Design Using Reference System Output y_m(t)}, using the
reference system output $y_m(t)$.

\subsection{Simulation System}
The linearized longitudinal dynamic model of an aircraft \cite[page 
138]{tctj04} has the form
\beq
\dot{x}_a(t) = A_a x_a(t) + b_a u_a(t) 
\label{pm7.1.1}
\eeq
where $x_a(t)=[U_b,W_b, Q_b, \theta_b]^T$ with $U_b$ being the aircraft
forward velocity ($x$-axis), $W_b$ being the vertical velocity
($z$-axis), $Q_b$ being the pitch rate ($y$-axis), and $\theta_b$ being
the Euler pitch angle ($y$-axis), $u_a(t)= dele$
is the system input (the angular position of the aircraft
elevator), and from \cite[page 155]{tctj04}, the parameters of (\ref{pm7.1.1}) are
\beq
A_a = \left[
\begin{array}{cccc}
-0.026373 &  0.12687 &  -12.926 &  -32.169 \\
-0.25009 &  -0.80174 &   220.55 &  -0.16307 \\
  0.000171 & -0.00754 &  -0.5510 &  -0.000334 \\
  0 &  0 &   1 &   0\\
\end{array}
\right],\;b_a = \left[
\begin{array}{c}
0.010887 \\
-0.18577 \\
-0.022966 \\
0 
\end{array}
\right].
\label{pm7.4.3}
\eeq
The output is chosen as the Euler pitch angle $\theta_b$ so that $c_a = 
\left[
\begin{array}{cccc}
0 & 0 & 0 & 1
\end{array}
\right]$. 

\bigskip
{\bf The plant (follower)}. For (\ref{pm7.1.1}) as the dynamic model of the plant (\ref{5x1}): 
$\dot{x}(t) = A x(t) + b u(t),\;y(t) = c x(t)$, with 
$A = A_a, b = b_a, c = c_a$, its transfer function is 
\beq
G(s)= c(sI - A)^{-1} b =
\frac{-0.023s^2-0.0176s-0.012}{s^4+1.3791s^3+2.1744s^2+0.989s+0.0651},
\label{5G(s)}
\eeq
whose relative degree is $n^* = 2$, and whose zeros are: $-0.6943$ and
$-0.0727$, both stable, and poles are: $-0.6760 \pm 1.2846j$ and
$-0.0136 \pm 0.1752j$, all stable but $-0.0136 \pm 0.1752j$ are not desired.

\bigskip
{\bf The reference model system (leader)}. To build a reference model
system (\ref{5x0}): $\dot{x}_{m}(t) = A_{m} x_{m}(t) + b_{m}
u_m(t),\;y_m(t) = c_m x_m(t)$, we consider a leader aircraft whose
dynamic model is also (\ref{pm7.1.1}), with $A_m = A_a, b_m = b_a, c_m
= c_a$, such that $G_m(s) = c_m(sI - A_m)^{-1} b_m = G(s)$ in
(\ref{5G(s)}). Since at least the reference system poles $-0.0136 \pm 0.1752j$ are not
desirable, we assume that a feedback control law is applied to improve
the reference system:
\beq
u_m(t) = k_{1m}^T x_m(t) + v_m(t),
\label{5rmsfcl}
\eeq
where $v_m(t)$ is a reference input signal, and $k_{1m} \in R^4$ is
from an LQR design:
\beq
k_{1m}^T = \left[
\begin{array}{cccc}
0.003614 &    -0.306976 &      262.056954   &    999.941914
\end{array}
\right]
\eeq
such that the closed-loop system matrix is
\beq
A_{mc} = A_m+b_m k_{1m}^T = 
\left[
\begin{array}{cccc}
-0.026334 &  0.123528 & -10.072986 & -21.282632 \\
-0.250761 & -0.744713 & 171.867680 & -185.922279 \\
 0.000088 & -0.000490 & -6.569400 & -22.965000\\
 0 &         0 &         1 &           0
\end{array}
\right]
\eeq
whose eigenvalues are: $-0.0732$, $-0.6897$, and $-3.2888 \pm
3.5045j$. We can see that the two eigenvalues $-0.0732$ and $-0.6897$ of
$A_{mc}$ are close to the system zeros $-0.6943$ and $-0.0727$, and the essential
eigenvalues $-3.2888 \pm 3.5045j$ are good ones, as they meet certain 
LQR optimality, so that the closed-loop reference system is good.

For adaptive control of the follower system (\ref{5x1}), the reference
system parameters $(A_m, b_m, c_m)$ are unknown, and so is its
feedback control law (\ref{5rmsfcl}), to the adaptive control law for 
(\ref{5x1}), and only the signals $y_m(t)$ (and $x_m(t)$) and
$u_m(t)$ are available, while the reference input signal $v_m(t)$ is
chosen and applied for some specified desired system behavior.

\medskip
\medskip
{\bf Nominal parameters}. For the plant (\ref{5G(s)}), the parameters satisfying (\ref{5732A.951}) are
\bea
\theta_1^* \ts = \ts [7.654562, 6.881398, -1.386117]^T,\nn\\
\theta_2^* \ts = \ts [-4434.189351, -5958.822119,
  -2550.603134]^T,\nn\\
\theta_{20}^* \ts = \ts 595.409427,\;\theta_3^* = -43.478261.
\eea

For the reference system (\ref{5x0}): $\dot{x}_{m}(t) = A_{m} x_{m}(t) + b_{m}
u_m(t),\;y_m(t) = c_m x_m(t)$, with $A_m = A_a, b_m = b_a, c_m
= c_a$, the parameter $\alpha_2 = -0.022966$ for (\ref{5r02u}) is calculated from
(\ref{5alpha2}), the parameters $\beta_i$, $i=1,2,20$, are defined in
(\ref{5r02}) with $\alpha_1 = [0.000171, -0.00754, 1.449, 0.999666]^T$ 
from (\ref{5alpha1}), for the matching
parameters $\theta_{3i}^*$, $i=1,2,3,4$, in (\ref{5thetas3i}), and then
for $\theta^*$ in (\ref{5thetasu}). 

\medskip
To detail the reduced-order observer formulation (\ref{5hatxm}), for
this reference system $(A_m, b_m, c_m)$, we choose $P_m = \left[
\begin{array}{c}
c_m\\
P_2
\end{array}
\right] \in R^{4 \times 4}$ with $P_2 = [I_3, 0_{3 \times 1}] \in R^{3
  \times 4}$, and form $\bar{A} = P_m A_m P_m^{-1} = 
\left[ \begin{array}{cc} \bar{A}_{11} & \bar{A}_{12}\\\bar{A}_{21} & \bar{A}_{22} 
\end{array} \right]$ and $\bar{B} = P_m B_m = 
\left[ \begin{array}{c} \bar{B}_1\\ \bar{B}_2 \end{array} \right]$,
where $\bar{A}_{11} \in R^{1 \times 1}$, 
$\bar{A}_{12} \in R^{1 \times 3}$, 
$\bar{A}_{21} \in R^{3 \times 1}$, 
$\bar{A}_{22} \in R^{3 \times 3}$, 
$\bar{B}_{1} \in R^{1 \times 1}$ and 
$\bar{B}_{2} \in R^{3 \times 1}$, which can be calculated as
\beq
\bar{A} = \left[
\begin{array}{cccc}
0 & 0 & 0 & 1 \\
-32.169  & -0.026373 & 0.12687 & -12.926 \\
-0.16307 & -0.25009 & -0.80174 & 220.55 \\
-0.000334 & 0.000171 & -0.00754 & -0.551
\end{array}
\right],\;
\bar{B} = \left[
\begin{array}{c}
                         0\\
                  0.010887\\
                  -0.18577\\
                 -0.022966
\end{array}
\right].
\eeq
Then, for $L_r \in R^{3 \times 1}$: $L_r = [3725.247701, -711.146582,
  4.620887]^T$,
 such that $\det(sI - (\bar{A}_{22} - L_r \bar{A}_{12})) = \Lambda_e(s)
= (s+2)^3$ chosen, the signal $w_m(t) \in
R^3$ is generated from the dynamic equation
\bea
\dot{w}_m(t) = (\bar{A}_{22} - L_r \bar{A}_{12}) w_m(t) + (\bar{B}_2 - L_r
\bar{B}_1) u_m(t) 
 + \;((\bar{A}_{22} - L_r \bar{A}_{12}) L_r + \bar{A}_{21}
- L_r \bar{A}_{11}) y_m(t),
\label{w(t)}
\eea
where $w_m(0) = w_0 \in R^{3}$ is an estimate of $\bar{x}_2(0)-L_r y_m(0)$
for $\bar{x}(t) = P_m x_m(t) = \left[ \begin{array}{c}
    \bar{x}_1(t)\\ \bar{x}_2(t) \end{array} \right]$ with 
$\bar{x}_1(t) = y_m(t)$ known and $\bar{x}_2(t) \in R^3$ to be
estimated. 

From (\ref{w(t)}), we can obtain (\ref{5wm(t)}): 
$w_m(t) = \frac{G_{u_m}(s)}{\Lambda_e(s)}[u_m](t) +
\frac{G_{y_m}(s)}{\Lambda_e(s)}[y_m](t) \in R^{3}$, with
\beq
\frac{G_{u_m}(s)}{\Lambda_e(s)} = \big(sI - (\bar{A}_{22} - L_r \bar{A}_{12})\big)^{-1}(\bar{B}_2 - L_r
\bar{B}_1) = \Theta_1^{*T} \frac{a(s)}{\Lambda_e(s)}
\eeq
\beq
\frac{G_{y_m}(s)}{\Lambda_e(s)} = \big(sI - (\bar{A}_{22} - L_r
\bar{A}_{12})\big)^{-1}((\bar{A}_{22} - L_r \bar{A}_{12}) L_r +
\bar{A}_{21} - L_r \bar{A}_{11}) = \Theta_2^{*T} \frac{a(s)}{\Lambda_e(s)}
\eeq
for some constant matrices $\Theta_1^* \in R^{3 \times 3}$ and
$\Theta_2^* \in R^{3 \times 3}$, which can be calculated as
\beq
\Theta_1^* = \left[
\begin{array}{ccc}
 60.879066 &  85.892363 &  0.010887 \\
-22.191198 & -22.365747 & -0.18577  \\
 -0.001159 &  -0.017616 & -0.022966
\end{array}
\right]^T
\eeq
\beq
\Theta_2^* = 
\left[
\begin{array}{ccc}
-30165.099352 & -37090.972425 & -17494.316383 \\       
  5725.209131 &   8197.711372 &   3943.609017 \\       
   -37.032155 &    -47.549544 &    -17.899977
\end{array}
\right],
\eeq
and $a(s) = [1, s, s^2]^T$. This expression of $w_m(t)$ is used
in (\ref{5hatxm}) for $\hat{x}_m(t)$, with $Q_m = P_m^{-1} = \left[
\begin{array}{cccc}
0_{3 \times 1} & & I_3 &\\
1 & 0 & 0 & 0
\end{array}
\right] \in R^{4 \times 4}$ so that $\hat{x}_m(t) = \left[\begin{array}{c}
w_m(t) + L_r y_m(t)\\
y_m(t)
\end{array}
\right]$. 

Then, from 
$r_m(t) = \alpha_1^T \hat{x}_m(t) + \alpha_2 u_m(t)$, we can obtain the parameters 
$\beta_1 \in R^{3}$, $\beta_{2} \in R^{3}$ and $\beta_{20} \in R$ for
(\ref{5r02}): 
$r_m(t) = \beta_1^T \omega_{u_m}(t) + \beta_{2}^T \omega_{y_m}(t) +
\beta_{20} y_m(t) + \alpha_2 u_m(t)$, where $\omega_{u_m}(t) = \frac{a(s)}{\Lambda_e(s)}
[u_m](t)$ and $\omega_{y_m}(t) = \frac{a(s)}{\Lambda_e(s)}[y_m](t)$. 
For $\alpha_1 = [\alpha_{11}, \alpha_{12}, \alpha_{13},
  \alpha_{14}]^T \stackrel{\triangle} = [\alpha_{0}^T, \alpha_{14}]^T$, we have
\bea
\alpha_1^T \hat{x}_m(t) \ts = \ts \alpha_0^T (w_m(t) + L_r y_m(t)) +
\alpha_{14} y_m(t) \nn\\
\ts = \ts \alpha_0^T \Theta_1^{*T} \omega_{u_m}(t) + 
\alpha_0^T \Theta_2^{*T} \omega_{y_m}(t) + 
\alpha_0^T L_r y_m(t) + \alpha_{14} y_m(t),
\eea
from which, we identify
\beq
\beta_1^T = \alpha_0^T \Theta_1^{*T},\;
\beta_2^T = \alpha_0^T \Theta_2^{*T},\;
\beta_{20} = \alpha_0^T L_r + \alpha_{14}.
\eeq

Finally, from (\ref{5thetas3i}) and (\ref{5thetasu}), we can obtain the 
parameter vector $\theta^*$ for the nominal control law
(\ref{5732A.932}): $u(t) = \theta^{*T} \omega(t)$, which is calculated
as
\bea
\theta^* \ts = \ts 
[\theta_{1}^{*T}, \theta_{2}^{*T}, \theta_{20}^*, 
\theta_{3}^* \beta_1^T, 
\theta_{3}^* \beta_2^T,\;
\theta_{3}^* \beta_{20}, 
\theta_{3}^* \alpha_2]^T \in R^{15} \nn\\
\ts = \ts [7.654562,
                  6.881398,
                 -1.386117,
              -4434.189351,
              -5958.822119,
\nn\\ \ts \ts 
              -2550.603134,
                595.409427, 
                 -7.654459,
                 -6.860858,
                  1.385877,
\nn\\ \ts \ts         
               4434.169641,
               5958.808248,
               2550.582910,
               -595.408430,
                  0.998522]^T.
\eea

For the adaptive control point of view, such a parameter
vector $\theta^*$ is unknown as the plant and reference system
parameters are unknown. In practice, the components of $\theta^*$ may
be known to be in certain ranges if the plant and reference system
parameters are known to be in some ranges. For simulation, the
knowledge of $\theta^*$ may be used to choose some reasonable values
of the initial parameter estimate $\theta(0)$, for
example, $\theta(0) = 0.95 \theta^*$ or $\theta(0) = 1.05 \theta^*$, 
to test the ability of the adaptive controller to handle the parameter
uncertainties.

\subsection{Simulation Results}
This study simulates the adaptive output tracking control system with
the controlled plant described by 
$\dot{x}(t) = A x(t) + b u(t),\;y(t) = c x(t)$, for 
$A = A_a$ and $b = b_a$ given in (\ref{pm7.4.3}) and $c = c_a
=[0,0,0,1]$ (the output $y(t)$ being the Euler pitch angle $\theta_b$) (the
follower aircraft). The 
reference system is described by $\dot{x}_{m}(t) = A_{m} x_{m}(t) + b_{m}
u_m(t),\;y_m(t) = c_m x_m(t)$, whose parameters are also $A_m = A_a,
b_m = b_a$ and $c_m = c_a$ (the leader aircraft) with a predesigned
feedback control law (\ref{5rmsfcl}): 
$u_m(t) = k_{1m}^T x_m(t) + v_m(t)$ (where $v_m(t)$ is a chosen
reference input), but unknown to the adaptive
controller for the plant, while the reference system input $u_m(t)$ (the
lead aircraft elevator angle) and output $y_m(t)$ (the lead aircraft
pitch angle) are known to the follower controller.

\medskip
In other words, this simulation is about a scenario in which the
 adaptive controller generating the elevator angle
$u(t)$ of the follower aircraft, with the only information of the
leader aircraft output $y_m(t)$ (pitch angle) and input $u_m(t)$
(elevator angle) (not that of $\dot{y}_m(t)$ and $\ddot{y}_m(t)$), can
ensure the follower aircraft output $y(t)$ (pitch angle) to track
$y_m(t)$ asymptotically, despite the uncertainties of $(A, b)$ and
$(A_m, b_m)$ (with $c = c_m = c_a = [0, 0, 0, 1]$ known in this case).

\medskip
For simulation, we consider the output feedback adaptive control scheme using the
reference system output $y_m(t)$, developed in Section \ref{Design
  Using Reference System Output y_m(t)}, whose controller structure is
\beq
u(t) = \theta_{1}^{T}\omega_{1}(t) + \theta_{2}^{T}\omega_{2}(t) + \theta_{20}
 y(t)  +  \theta_{31}^{T} \omega_{u_m}(t) + \theta_{32}^{T} \omega_{y_m}(t) +
\theta_{33} y_m(t) + \theta_{34} u_m(t),
\label{5simucl}
\eeq
where $\theta_{1} \in R^3$, $\theta_{2} \in R^3$, $\theta_{20} \in R$,
$\theta_{31} \in R^3$, $\theta_{32} \in R^3$, $\theta_{33} \in R$,
$\theta_{34} \in R$, and
\begin{equation}
\omega_{1}(t) = \frac{a(s)}{\Lambda(s)}[u](t),\;\omega_{2}(t) = 
\frac{a(s)}{\Lambda(s)}[y](t)
\end{equation}
\beq
\omega_{u_m}(t) = \frac{a(s)}{\Lambda_e(s)}[u_m](t),\;\omega_{y_m}(t) = \frac{a(s)}{\Lambda_e(s)}[y_m](t),
\eeq
for $a(s) = [1, s, s^2]^{T}$ with $\Lambda_e(s) = \Lambda(s) =
(s+2)^3$ chosen. 

The virtual reference model system (\ref{5mrs}) is: $y_m(t)
= W_m(s)[r_m](t)$, where $W_m(s) = \frac{1}{(s+1)^2}$ is chosen and
$y_m(t)$ is from the actual reference system (\ref{5x0}):
$\dot{x}_{m}(t) = A_{m} x_{m}(t) + b_{m} u_m(t),\;y_m(t) = c_m
x_m(t)$, with $u_m(t)$ generated from the control law
(\ref{5rmsfcl}). The parameters of the control law (\ref{5simucl}) are
updated from the adaptive laws (\ref{5uthetalaw})-(\ref{5urholaw})
with $\sign[k_p] = -$ for $k_p = -0.023$ in (\ref{5G(s)}), 
adaptation gains $\Gamma = 5 I_{15}$ and $\gamma = 5$, and initial
values $\theta(0) = 1.1 \theta^*$ and $\rho(0) = 1.1 \rho^*$.

\medskip
Some typical simulation results, demonstrating the desired (and analytically
ensured) adaptive control system performance, for $v_m(t) = 300$ are
shown in Figure 1: the tracking error $e(t) = y(t) - y_m(t)$ and the
plant output $y(t)$ versus the reference system output $y_m(t)$ (whose
steady-state value is approximately equal to $-0.3$ radians or
$-17.1887 = - 0.3 \times 57.2958$ degrees), and the results for
$v_m(t) = 300 \sin(0.3t)$ are shown in Figure 2 for $e(t)$ and for $y(t)$
versus $y_m(t)$. Some additional simulation results include that 
for $v_m(t) = 300 \sin(0.3t) + 250 \sin(0.5t)$ shown in Figure 3 for
$e(t)$ and for $y(t)$ versus $y_m(t)$, and that for 
$v_m(t) = 300 \sin(0.3t) + 250 \sin(0.5t) + 200 \sin(0.7t)$ shown in
Figure 4 for $e(t)$ and for $y(t)$ versus $y_m(t)$, simulating
different pitch angle tracking trajectories.

\begin{figure}[p]
\label{fig1}
\centering
\includegraphics[height=4in,width=5.0in,angle=0]{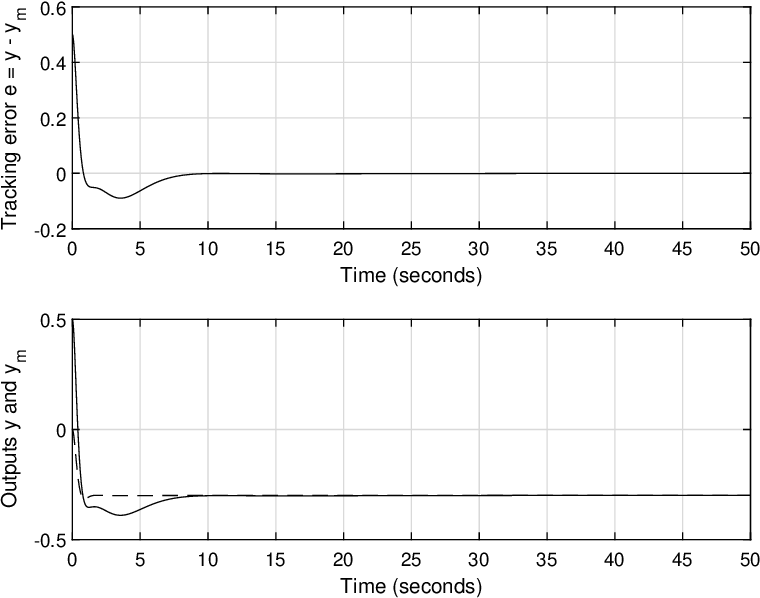}
\caption{System responses for $v_m(t) = 300$.}
\bigskip
\bigskip
\label{fig2}
\centering
\includegraphics[height=4in,width=5.0in,angle=0]{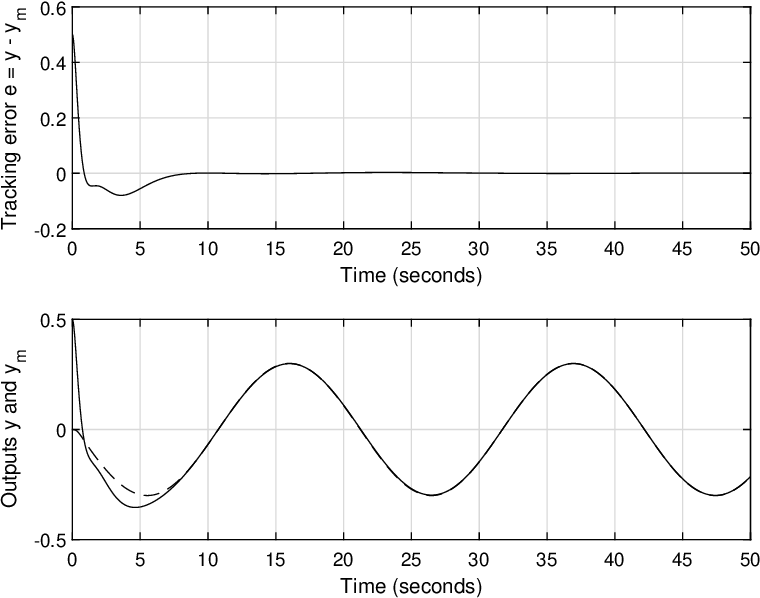}
\caption{System responses for $v_m(t) = 300 \sin (0.3t)$.}
\end{figure}

\begin{figure}[p]
\label{fig3}
\centering
\includegraphics[height=4in,width=5.0in,angle=0]{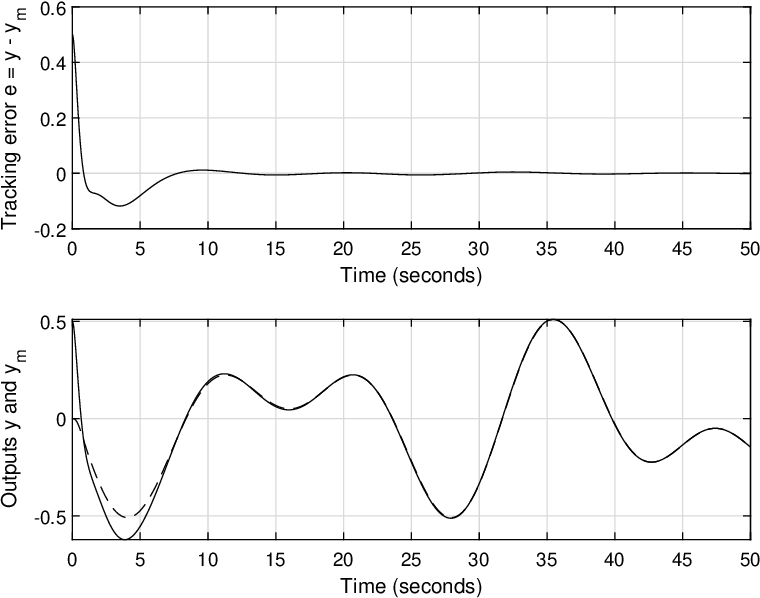}
\caption{System responses for $v_m(t) = 300 \sin (0.3t) + 250 \sin (0.5t)$.}
\bigskip
\bigskip
\label{fig4}
\centering
\includegraphics[height=4in,width=5.0in,angle=0]{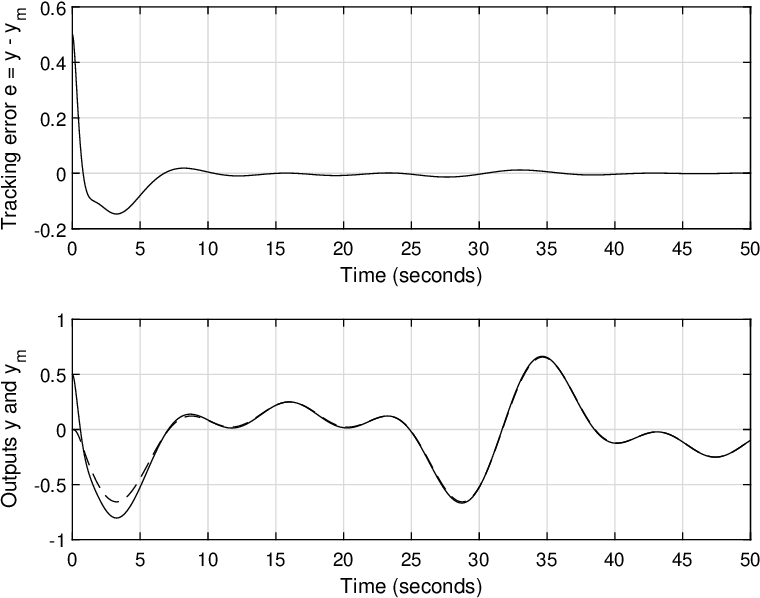}
\caption{System responses for $v_m(t) = 300 \sin (0.3t) + 250 \sin
  (0.5t) + 200 \sin(0.7t)$.}
\end{figure}

\section{Concluding Remarks}
In this paper, we have developed new schemes to expand the capacity of
model reference adaptive control to deal with the uncertainties of the
reference model system which cannot meet the conditions of a standard
reference system in traditional model reference adaptive control which
requires the knowledge of either reference output derivatives or
reference system dynamics. A key feature of the developed new schemes
is the controller structure expansion with parametrized estimation
of the reference output derivatives and reference system
parameters. The adaptive output tracking control objective is achieved
with state feedback and output feedback control designs using the
input signal plus either the state variable signals or the output
signal of the reference system. Desired stability and tracking
properties of the expanded adaptive control systems are established 
 in theory and illustrated by simulation results. Similar studies have
 indicated that extensions to adaptive tracking control of other
 systems are also possible, using the developed controller structure
 expansion method, to deal with the uncertainties of the reference
 model system, which are not present in the traditional model
 reference adaptive control problems.

\end{document}